\documentclass[a4paper,12p,reqno]{amsart}   

\usepackage{ifthen}
\ifthenelse{\isundefined \OlafsVersion}
{

\newcommand{\myfont}{\sffamily}

\newtheoremstyle{mythmstyle}
  {\topsep}
  {\topsep}
  {\itshape}
  {}
  {\bfseries \myfont}
  {.}
  {.5em}
  {}

\newtheoremstyle{mydefstyle}
  {\topsep}
  {\topsep}
  {\normalfont}
  {}
  {\bfseries \myfont}
  {.}
  {.5em}
  {}

\swapnumbers                    

\theoremstyle{mythmstyle}       

\let\MakeUppercase\relax

\expandafter\let\expandafter\oldproof\csname\string\proof\endcsname
\let\oldendproof\endproof
\renewenvironment{proof}[1][\bfseries\myfont\proofname]{%
  \oldproof[\bfseries \myfont #1]%
}{\oldendproof}

\makeatletter
\newtoks\thm@headfont  \thm@headfont{\bfseries \myfont}

\renewcommand\section{\@startsection{section}{1}%
  \z@{.7\linespacing\@plus\linespacing}{.5\linespacing}%
  {\Large\myfont\bfseries}}
\renewcommand\subsection{\@startsection{subsection}{2}%
  \z@{-.5\linespacing\@plus-.7\linespacing}{.5\linespacing}%
  {\large\myfont\bfseries}}
\renewcommand\subsubsection{\@startsection{subsubsection}{3}%
  \z@{.5\linespacing\@plus.7\linespacing}{-.5em}%
  {\myfont\bfseries}}
\renewenvironment{abstract}{%
  \ifx\maketitle\relax
    \ClassWarning{\@classname}{Abstract should precede
      \protect\maketitle\space in AMS document classes; reported}%
  \fi
  \global\setbox\abstractbox=\vtop \bgroup
    \normalfont\Small
    \list{}{\labelwidth\z@
      \leftmargin3pc \rightmargin\leftmargin
      \listparindent\normalparindent \itemindent\z@
      \parsep\z@ \@plus\p@
      
    }%
    \item[\hskip\labelsep
      \myfont\bfseries
    \abstractname.]%
}{%
  \endlist\egroup
  \ifx\@setabstract\relax \@setabstracta \fi
}
\renewcommand\contentsnamefont{\myfont\bfseries}
\renewcommand\@starttoc[2]{\begingroup
  \setTrue{#1}%
  \par\removelastskip\vskip\z@skip
  \@startsection{}\@M\z@{\linespacing\@plus\linespacing}%
    {.5\linespacing}{
      \contentsnamefont}{#2}%
  \ifx\contentsname#2%
  \else \addcontentsline{toc}{section}{#2}\fi
  \makeatletter
  \@input{\jobname.#1}%
  \if@filesw
    \@xp\newwrite\csname tf@#1\endcsname
    \immediate\@xp\openout\csname tf@#1\endcsname \jobname.#1\relax
  \fi
  \global\@nobreakfalse \endgroup
  \addvspace{32\p@\@plus14\p@}%
  \let\tableofcontents\relax
}
\renewcommand\@settitle{\begin{center}%
  \baselineskip14\p@\relax
    \LARGE
    \bfseries
    \myfont
  \@title
  \end{center}%
}
\renewcommand\@setauthors{%
  \begingroup
  \def\thanks{\protect\thanks@warning}%
  \trivlist
  \centering\footnotesize \@topsep30\p@\relax
  \advance\@topsep by -\baselineskip
  \item\relax
  \author@andify\authors
  \def\\{\protect\linebreak}%
  \large
  \myfont\bfseries\authors
  \ifx\@empty\contribs
  \else
    ,\penalty-3 \space \@setcontribs
    \@closetoccontribs
  \fi
  \endtrivlist
  \normalfont\myfont\@setaddresses
  \endgroup
}
\renewcommand\@setaddresses{\par
  \nobreak \begingroup
\footnotesize
  \def\author##1{\nobreak\addvspace\bigskipamount}%
  \def\\{\unskip, \ignorespaces}%
  \interlinepenalty\@M
  \def\address##1##2{\begingroup
    \par\addvspace\bigskipamount\indent
    \@ifnotempty{##1}{(\ignorespaces##1\unskip) }%
    {
      \ignorespaces##2}\par\endgroup}%
  \def\curraddr##1##2{\begingroup
    \@ifnotempty{##2}{\nobreak\indent\curraddrname
      \@ifnotempty{##1}{, \ignorespaces##1\unskip}\/:\space
      ##2\par}\endgroup}%
  \def\email##1##2{\begingroup
    \@ifnotempty{##2}{\nobreak\indent\emailaddrname
      \@ifnotempty{##1}{, \ignorespaces##1\unskip}\/:\space
      \ttfamily##2\par}\endgroup}%
  \def\urladdr##1##2{\begingroup
    \def~{\char`\~}%
    \@ifnotempty{##2}{\nobreak\indent\urladdrname
      \@ifnotempty{##1}{, \ignorespaces##1\unskip}\/:\space
      \ttfamily##2\par}\endgroup}%
  \addresses
  \endgroup
}
\renewcommand\enddoc@text{\ifx\@empty\@translators \else\@settranslators\fi
}
\renewcommand\@secnumfont{\myfont\bfseries} 

\renewcommand\maketitle{\par
  \@topnum\z@ 
  \@setcopyright
  \thispagestyle{firstpage}
  \ifx\@empty\shortauthors \let\shortauthors\shorttitle
  \else \andify\shortauthors
  \fi
  \@maketitle@hook
  \begingroup
  \@maketitle
  \toks@\@xp{\shortauthors}\@temptokena\@xp{\shorttitle}%
  \toks4{\def\\{ \ignorespaces}}
  \edef\@tempa{%
    \@nx\markboth{\the\toks4
      \@nx\MakeUppercase{\the\toks@}}{\the\@temptokena}}%
  \@tempa
  \endgroup
  \c@footnote\z@
  \@cleartopmattertags
}
\def\@captionheadfont{\myfont\bfseries} 

\makeatother
\usepackage{amsthm}

\swapnumbers  
\newcounter{intro}

\newtheorem{maintheorem}[intro]{Theorem}
\newtheorem{maincorollary}[intro]{Corollary}

\swapnumbers 
\newtheorem{theorem}{Theorem}[section]
\newtheorem{proposition}[theorem]{Proposition}
\newtheorem{lemma}[theorem]{Lemma}
\newtheorem{corollary}[theorem]{Corollary}

\theoremstyle{mydefstyle}        
\newtheorem{definition}[theorem]{Definition}

\newtheorem{remark}[theorem]{Remark}

\newtheorem*{remark*}{Remark}

\numberwithin{equation}{section}
\newcommand{\Sec}[1]{Section~\ref{sec:#1}}

\newcommand{\App}[1]{Appendix~\ref{app:#1}}

\newcommand{\Subsec}[1]{Subsection~\ref{ssec:#1}}

\newcommand{\Fig}[1]{Figure~\ref{fig:#1}}

\newcommand{\Figs}[2]{Figures~\ref{fig:#1} and~\ref{fig:#2}}

\newcommand{\Thm}[1]{Theorem~\ref{thm:#1}}
\newcommand{\Thms}[2]{Theorems~\ref{thm:#1} and~\ref{thm:#2}}

\newcommand{\Lem}[1]{Lemma~\ref{lem:#1}}
\newcommand{\Lems}[2]{Lemmata~\ref{lem:#1} and~\ref{lem:#2}}
\newcommand{\LemS}[2]{Lemmata~\ref{lem:#1}--\ref{lem:#2}}

\newcommand{\Cor}[1]{Corollary~\ref{cor:#1}}

\newcommand{\Prp}[1]{Proposition~\ref{prp:#1}}

\newcommand{\Prpenum}[2]{Proposition~\ref{prp:#1}~(\ref{#2})}

\newcommand{\Def}[1]{Definition~\ref{def:#1}}
\newcommand{\Defs}[2]{Definitions~\ref{def:#1} and~\ref{def:#2}}
\newcommand{\Defss}[3]{Definitions~\ref{def:#1},~\ref{def:#2} and~\ref{def:#3}}

\newcommand{\Defenum}[2]{Definition~\ref{def:#1}~(\ref{#2})}

\usepackage{accents}
\newcommand{\abs}[2][{}]{\lvert{#2}\rvert_{{#1}}}    
\newcommand{\abssqr}[2][{}]{\lvert{#2}\rvert^2_{#1}} 
\newcommand{\bigabs}[2][{}]{\bigl\lvert{#2}\bigr\rvert_{#1}}     
\newcommand{\bigabssqr}[2][{}]{\bigl\lvert{#2}\bigr\rvert^2_{#1}}
\newcommand{\Bigabssqr}[2][{}]{\Bigl\lvert{#2}\Bigr\rvert^2_{#1}}

\newcommand{\normsymb}{\|}
\newcommand{\bignormsymb}[1]{#1\|}

\newcommand{\norm}[2][{}]{\normsymb{#2}\normsymb_{{#1}}}    
\newcommand{\normsqr}[2][{}]{\normsymb{#2}\normsymb^2_{#1}} 

\newcommand{\bignormsqr}[2][{}]{\bignormsymb{\bigl}{#2}%
                                \bignormsymb{\bigr}^2_{#1}}

\newcommand{\iprod}[3][{}]{\langle{#2},{#3}\rangle_{#1}}  

\newcommand{\set}[2]{\{ \, #1 \, | \, #2 \, \} }      
\newcommand{\bigset}[2]{\bigl\{ \, #1 \, \bigl|\bigr. \, #2 \, \bigr\} }

\DeclareMathOperator*{\dcup}   {\mathaccent\cdot\cup}

\newcommand{\map}[3]{ #1 \colon #2 \longrightarrow #3}    
\newcommand{\embmap}[3]{ #1 \colon #2 \hookrightarrow #3} 

\newcommand{\bd}  {\partial}          
\newcommand{\clo}[2][]{\overline{{#2}}^{#1}} 
\newcommand{\intr}[1]{\ring{{#1}}}    

\newcommand{\restr}[1]{{\restriction}_{#1}} 

\def\Xint#1{\mathchoice
   {\XXint\displaystyle\textstyle{#1}}%
   {\XXint\textstyle\scriptstyle{#1}}%
   {\XXint\scriptstyle\scriptscriptstyle{#1}}%
   {\XXint\scriptscriptstyle\scriptscriptstyle{#1}}%
   \!\int}
\def\XXint#1#2#3{{\setbox0=\hbox{$#1{#2#3}{\int}$}
     \vcenter{\hbox{$#2#3$}}\kern-.5\wd0}}
\def\XXsum#1#2#3{{\setbox0=\hbox{$#1{#2#3}{\int}$}
     \vcenter{\hbox{$#2#3$}}\kern-.60\wd0}}

\newcommand{\dashint}{\Xint-}   

\newcommand{\card}[1]{\lvert#1\rvert}   

\newcommand{\dd}    {\, \mathrm d}    

\DeclareMathOperator{\dom}    {dom}
\DeclareMathOperator{\ran}    {ran}
\DeclareMathOperator{\id}     {id}   
\DeclareMathOperator{\ind}    {ind}  

\DeclareMathOperator{\vol}    {vol}

\newcommand{\specsymb} {\sigma} 

\newcommand{\spec}[2][{}]   {\specsymb_{\mathrm{#1}}(#2)}

\newcommand{\eps}{\varepsilon} 

\renewcommand{\phi}{\varphi}   
\renewcommand{\rho}{\varrho}   

\DeclareMathOperator{\myRe} {Re}
\renewcommand{\Re}     {\myRe}

\newcommand{\conj}[1]{\overline {#1}}

\newcommand{\R}{\mathbb{R}} 
\newcommand{\C}{\mathbb{C}} 
\newcommand{\N}{\mathbb{N}} 

\newcommand{\1}{\mathbbm 1}                    

\newcommand{\im}{\mathrm i} 

\newcommand{\wt}{\widetilde}           

\newcommand{\Err}{\mathrm O}

\newcommand{\HS}{\mathscr H}           

\newcommand{\Sobsymb} {\mathsf H} 

\newcommand{\Contsymb} {\mathsf C}     
\newcommand{\Lsymb}    {\mathsf L}     
\newcommand{\lsymb}    {\ell}          

\newcommand{\Sobspace}[1][1]{\Sobsymb^{{#1}}}

\newcommand{\Contspace}[1][{}]{\Contsymb^{{#1}}}     
\newcommand{\Lpspace}[1][p]    {\Lsymb_{#1}}     
\newcommand{\lpspace}[1][p]    {\lsymb_{#1}}     
\newcommand{\Lsqrspace}    {\Lpspace[2]}     
\newcommand{\lsqrspace}    {\lpspace[2]}          

\newcommand{\Ci} [2][{}]{\Contspace [\infty]_{#1} ({#2})}
\newcommand{\Cci}[1]{\Ci[\mathrm c]{#1}}
\newcommand{\Cont}[2][{}]{\Contspace[#1]({#2})}

\newcommand{\Lsqr}[2][{}]{\Lsqrspace^{#1}({#2})} 
\newcommand{\lsqr}[2][{}]{\lsqrspace^{#1}({#2})}   

\newcommand{\Sob}[2][1]{\Sobspace [{#1}]({#2})}         

\newcommand{\SobWsymb}{\mathsf W} 

\newcommand{\SobWspace}[2][p]{\SobWsymb_{#1}^{#2}}

\newcommand{\Sobx}[3][1]{\Sobspace [#1]_{{#2}}({#3})} 

\newcounter{myenumi}

\newcommand{\itemref}[1]{\eqref{#1}}

\newcommand{\quadtext}[1]{\quad\text{#1}\quad}
\newcommand{\qquadtext}[1]{\qquad\text{#1}\qquad}

}
{
  \input{\OlafsPath/olaf-ams-layout-macros.tex}
  \input{\OlafsPath/olaf-thm-environment-macros.tex}
  \input{\OlafsPath/olaf-refs-unified.tex}
  \input{\OlafsPath/olaf-math-symbols.tex}
  \input{\OlafsPath/olaf-spaces-macros.tex}
  \input{\OlafsPath/olaf-other-macros-myenumi-myparagraph.tex}
}

\usepackage[textwidth=2.6cm,backgroundcolor=black!10,textsize=tiny,%
colorinlistoftodos]{todonotes}

\ifthenelse{\isundefined \DraftVersion}
{ 
  \newcommand{\PEtodo}[1]{}
  \newcommand{\OPtodo}[1]{}
  
}
{ 
  \usepackage[norefs,nocites]{refcheck}     

  \usepackage[notref,notcite]{showkeys}     

  \newcommand{\PEtodo}[1]{\todo[fancyline,backgroundcolor=blue!10]{PE: #1}} 
  \newcommand{\OPtodo}[1]{\todo[fancyline,backgroundcolor=black!10]{OP: #1}} 
}

  \usepackage{datetime}    
  
  \yyyymmdddate   
  \AddToHook{shipout/background}[mypic]
  {
    \put(2.5cm,-28.5cm)%
    {\footnotesize \today, \currenttime, \emph{File:} \texttt{\jobname.tex}}

  }
\usepackage[a4paper]{geometry}
\usepackage{amsmath,amsxtra,amssymb,amsthm}
\usepackage{mathrsfs}  
\usepackage{bbm}

\usepackage{tikz}
\usepackage{graphicx}

\usepackage{hyperref}

\usepackage{bm}            
\usepackage{interval}      

\newcommand{\Xzero}[1][]{%
  \ifthenelse{\equal{#1}{}}%
  {X_0} {X_{0,#1}}%
}
\newcommand{\intrXzero}[1][e]{%
  \ifthenelse{\equal{#1}{}}%
  {\intr X_0} {\intr X_{0,#1}}%
}  

\newcommand{\X}{X}             
\newcommand{\wtX}{\widetilde X}             
\newcommand{\Xeps}{\X_\eps}    
\newcommand{\Xepsed}[1][]{%
  \ifthenelse{\equal{#1}{}}%
  {\X_{\eps,e}}%
  {\X_{\eps,e#1}}%
}    
\newcommand{\Xepsvx}[1][]{
  \ifthenelse{\equal{#1}{}}%
  {\X_{\eps,v}}%
  {\X_{\eps,v#1}}%
}    
\newcommand{\wtXone}{\wtX_1}    
\newcommand{\wtXoneed}[1][]{%
  \ifthenelse{\equal{#1}{}}%
  {\wtX_{1,e}}%
  {\wtX_{1,e,#1}}%
}    
\newcommand{\wtXonevx}[1][]{
  \ifthenelse{\equal{#1}{}}%
  {\wtX_{1,v}}%
  {\wtX_{1,v,#1}}%
}    
\newcommand{\wtXeps}{\wtX_\eps}    
\newcommand{\wtXepsed}[1][]{%
  \ifthenelse{\equal{#1}{}}%
  {\wtX_{\eps,e}}%
  {\wtX_{\eps,e,#1}}%
}    
\newcommand{\wtXepsvx}[1][]{
  \ifthenelse{\equal{#1}{}}%
  {\wtX_{\eps,v}}%
  {\wtX_{\eps,v,#1}}%
}    
\newcommand{\Xone}{\X_1}        
\newcommand{\Xoneed}[1][]{%
  \ifthenelse{\equal{#1}{}}%
  {\X_{1,e}}%
  {\X_{1,e,#1}}%
}    
\newcommand{\Xonevx}[1][]{
  \ifthenelse{\equal{#1}{}}%
  {\X_{1,v}}%
  {\X_{1,v,#1}}%
}    
\newcommand{\Y}{Y}             
\newcommand{\Yone}{\Y_1}        
\newcommand{\Yepsed}{\Y_{\eps,e}}       
\newcommand{\Yoneed}{\Y_{1,e}}        

\newcommand{\stSh}[1]{#1^\#}  
\newcommand{\starXepsvx}{\stSh \X_{\eps,v}}%

\newcommand{\stShgSxx}[2]   
{%
  \ifthenelse{\boolean{EpsNotation}}
  {#1^\#_{\eps,#2}} 
  {\widetilde{#1}^\#_{#2}}   
}  

\newcommand{\CGaffney}{C_{\mathrm{Gaffney}}}
\newcommand{\Cvx}{C_{\mathrm{vx}}}
\newcommand{\Cvxcol}{C_{\mathrm{vx\,col}}}
\newcommand{\Cisoper}{C_{\mathrm{isoper}}}
\newcommand{\lambdaVx}{\lambda_2^{\mathrm{vx}}}
\newcommand{\lambdaEd}{\lambda_2^{\mathrm{ed}}}
\newcommand{\curvMax}{\kappa_{\max}}

\newcommand{\dec}{\mathrm{dec}}
\newcommand{\Sobdec}[2][1]{\Sobx[#1] \dec {#2}}
\newcommand{\Sobsum}[2][1]{\Sobx[#1] \Sigma {#2}}

\newcommand{\normvec}[1]{\mathrm n_{#1}}  

\renewcommand{\vol}[2][]{\abs[#1]{#2}}  
\renewcommand{\id}[1]{\operatorname {id}_{#1}} 
\newcommand{\lebesgue}{\bm \lambda}  

\newcommand{\Lin}[1]{\mathscr L(#1)}

\newcommand{\intbd}{\ring \bd}
\DeclareMathOperator{\II}{II}  
\DeclareMathOperator{\dHaus}{d_{Hausd}}  

\def\Xint#1{\mathchoice
{\XXint\displaystyle\textstyle{#1}}%
{\XXint\textstyle\scriptstyle{#1}}%
{\XXint\scriptstyle\scriptscriptstyle{#1}}%
{\XXint\scriptscriptstyle\scriptscriptstyle{#1}}%
\!\int}
\def\XXint#1#2#3{{\setbox0=\hbox{$#1{#2#3}{\int}$ }
\vcenter{\hbox{$#2#3$ }}\kern-.6\wd0}}
\def\dashint{\Xint-}

\def\myXone#1{\mathchoice
{\XXone\displaystyle\textstyle{#1}}%
{\XXone\textstyle\scriptstyle{#1}}%
{\XXone\scriptstyle\scriptscriptstyle{#1}}%
{\XXone\scriptscriptstyle\scriptscriptstyle{#1}}%
\!\1}
\def\XXone#1#2#3{{\setbox0=\hbox{$#1{#2#3}{\1}$ }
\vcenter{\hbox{$#2#3$ }}\kern-.50\wd0}}
\newcommand{\dashOne}{\myXone-}

\newboolean{EpsNotation}
\setboolean{EpsNotation}{false}  

\newcommand{\mG}[1]{
  \ifthenelse{\boolean{EpsNotation}}
  {#1_0} 
  {#1}   
}  

\newcommand{\mGed}[2][e]
{
  \ifthenelse{\boolean{EpsNotation}}
  {{#2}_{0,#1}} 
  {#2_{#1}}   
}  

\newcommand{\intmGxx}[2]   
{%
  \ifthenelse{\boolean{EpsNotation}}
  {{\intr #1}_{0,#2}} 
  {\intr{#1}_{#2}}   
}  

\newcommand{\gS}[1]{
  \ifthenelse{\boolean{EpsNotation}}
  {#1_\eps} 
  {\widetilde{#1}}   
}  

\newcommand{\gSeps}[1]
{%
  \ifthenelse{\boolean{EpsNotation}}
  {\eps #1} 
  {\widetilde{#1}}   
}  
\newcommand{\gSepS}[1]  
{%
  \ifthenelse{\boolean{EpsNotation}}
  {#1} 
  {\widetilde {#1}}   
}  
\newcommand{\gSxx}[2]
{%
  \ifthenelse{\boolean{EpsNotation}}
  {\ifthenelse{\equal{#2}{}}
    {{#1}_\eps}
    {{#1}_{\eps,#2}
    }
  } 
  {\ifthenelse{\equal{#2}{}}
    {\widetilde{#1}}
    {\widetilde{#1}_{#2}}
  }   
}  

\newcommand{\gSed}[2][e] {\gSxx{#2}{#1}} 
\newcommand{\gSvx}[2][v] {\gSxx{#2}{#1}}  

\newcommand{\intgSxx}[2]   
{%
  \ifthenelse{\boolean{EpsNotation}}
  {{\intr #1}_{\eps,#2}} 
  {\intr{\widetilde{#1}}_{#2}}   
}  
\newcommand{\intgS}[1]  {\intgSxx{#1}{}} 

\newcommand{\FF} {F} 
\newcommand{\ff} {f} 
\newcommand{\Ff} {\vec f} 
\newcommand{\GG} {G} 

\newcommand{\UU} {U} 
\newcommand{\uu} {u} 
\newcommand{\Uu} {\vec u} 
\newcommand{\WW} {W} 
\newcommand{\ww} {w} 
\newcommand{\Ww} {\vec w} 
\newcommand{\tuu} {\phi} 
\newcommand{\tUu} {\vec \phi} 

\newcommand{\KK} {K} 
\newcommand{\LL} {L} 

\newcommand{\hs}{\HS^0}   
\newcommand{\Hs}{\HS^1}   
\newcommand{\HSdom}{\mathscr D}   
\newcommand{\hsdom}{\HSdom^0}   
\newcommand{\Hsdom}{\HSdom^1}   

\newcommand{\Hsdomeps}{\HSdom^1_\eps}   

\newcommand{\wtHsdomeps}{\wt \HSdom^1_\eps}   

\newcommand{\mHS}{\mG \HS}   
\newcommand{\mhs}{\mG \hs}   
\newcommand{\mHs}{\mG \Hs}   

\newcommand{\HSeps}{\HS_\eps}   

\newcommand{\wtHSeps}{\widetilde \HS_\eps}   
\newcommand{\wthseps}{\widetilde \HS^0_\eps}   
\newcommand{\wtHseps}{\widetilde \HS^1_\eps}   

\newcommand{\HSzero}{\HS_0}   

\newcommand{\Diraceps}{\Dirac_\eps}   
\newcommand{\Diraczero}{\Dirac_0}   

\newcommand{\mHSdom}{\mG \HSdom}   
\newcommand{\mhsdom}{\mG \hsdom}   
\newcommand{\mHsdom}{\mG \Hsdom}   

\newcommand{\xHS}{\gS \HS}   
\newcommand{\xhs}{\gS \hs}   
\newcommand{\xHs}{\gS \Hs}   

\newcommand{\xhsed}{\gSed \HS^0}   
\newcommand{\xHsed}{\gSed \HS^1}   
\newcommand{\xhsvx}{\gSvx \HS^0}   

\newcommand{\xHSdom}{\gS \HSdom}   
\newcommand{\xhsdom}{\gS \hsdom}   
\newcommand{\xHsdom}{\gS \Hsdom}   

\newcommand{\Dirac}{D}      
\newcommand{\Grad}{d}    
\newcommand{\wtGrad}{\widetilde d}    
\newcommand{\ExtDer}[1][]
{
  \ifthenelse{\equal{#1}{}}
  {d}{d^{#1}}
}
\newcommand{\GradAdj}{d^*}   
\newcommand{\wtGradAdj}{\widetilde d^*}   
\newcommand{\Div}[1][]
{
  \ifthenelse{\equal{#1}{}}
  {\delta}{\delta^{#1}}
}

\newcommand{\LAPL}{\Delta}   
\newcommand{\lapl}{\Delta^0}   
\newcommand{\Lapl}{\Delta^1}   

\newcommand{\mDirac}{\mG \Dirac}   
\newcommand{\mGrad}{\mG \Grad}   
\newcommand{\mGradAdj}{\mG \GradAdj}   

\newcommand{\mLAPL}{\mG \LAPL}   
\newcommand{\mlapl}{\mG \lapl}   
\newcommand{\mLapl}{\mG \Lapl}   

\newcommand{\transv}[1]{#1^\bot}
\newcommand{\longit}[1]{#1^\|}

\newcommand{\Forms}[3][]{\mathord{\textstyle\bigwedge}^{#2}
  T\ifthenelse{\equal{#1}{}}{}{_{#1}}^*#3}

\newcommand{\Tensor}[3][]%
{%
  \ifthenelse{\equal{#1}{!}}
  {(T^*#3)^{\otimes #2}}
  {\ifthenelse{\equal{#1}{}}
    {T^*#3^{\otimes #2}}
    {T_{#1}^*#3^{\otimes #2}}
  }
}

\newcommand{\xDirac}{\gS \Dirac}   
\newcommand{\wtxDirac}{\wt \Dirac_\eps}   
\newcommand{\xGrad}{\gS \Grad}   
\newcommand{\xGradAdj}{\gS \GradAdj}   

\newcommand{\xtransvGrad}{\transv{\gS \Grad}}   
\newcommand{\xtransvGradAdj}{(\transv{\gS \Grad})^*}   

\newcommand{\efct}[1][]{\gSxx {\dashOne}{#1} }

\newcommand{\xLAPL}{\gS \Delta}   
\newcommand{\xlapl}{\gS \Delta^0}   
\newcommand{\xLapl}{\gS \Delta^1}   

\newcommand{\hsaux}{\mathscr G^0}  
\newcommand{\Hsaux}{\mathscr G^1}  

\newcommand{\mOpA}{\mG A}
\newcommand{\mOpB}{\mG B}
\newcommand{\xOpA}{\gS A}
\newcommand{\xOpB}{\gS B}

\newcommand{\Res}{R}      
\newcommand{\A}{D}         
\newcommand{\xA}{\wt D}        

\newcommand{\mRes}{\mG \Res}      
\newcommand{\xRes}{\gS \Res}      

\newcommand{\Reseps}{R_\eps}      
\newcommand{\wtReseps}{\widetilde R_\eps}      

\newcommand{\IdOp}{
  \ifthenelse{\boolean{EpsNotation}}
  {J_\eps} 
  {J}   
} %
\newcommand{\idop}{
  \ifthenelse{\boolean{EpsNotation}}
  {J_\eps^0} 
  {J^0}   
} %
\DeclareRobustCommand*
{\Idop}{
  \ifthenelse{\boolean{EpsNotation}}
  {J_\eps^1} 
  {J^1}   
} %
\newcommand{\IdOpAdj}{
  \ifthenelse{\boolean{EpsNotation}}
  {J_\eps^*} 
  {J^*}   
} %
\newcommand{\idopAdj}{
  \ifthenelse{\boolean{EpsNotation}}
  {(J_\eps^0)^*} 
  {(J^0)^*}   
} %
\newcommand{\IdopAdj}{
  \ifthenelse{\boolean{EpsNotation}}
  {(J_\eps^1)^*} 
  {(J^1)^*}   
} %

\newcommand{\wtIdOp}{
  \ifthenelse{\boolean{EpsNotation}}
  {\wt J_\eps} 
  {\wt J}   
} %
\newcommand{\wtidop}{
  \ifthenelse{\boolean{EpsNotation}}
  {\wt J_\eps^0} 
  {\wt J^0}   
} %
\newcommand{\wtIdop}{
  \ifthenelse{\boolean{EpsNotation}}
  {\wt J_\eps^1} 
  {\wt J^1}   
} %
\newcommand{\wtIdOpAdj}{
  \ifthenelse{\boolean{EpsNotation}}
  {\wt J_\eps^*} 
  {\wt J^*}   
} %
\newcommand{\wtidopAdj}{
  \ifthenelse{\boolean{EpsNotation}}
  {(\wt J_\eps^0)^*} 
  {(\wt J^0)^*}   
} %
\newcommand{\wtIdopAdj}{
  \ifthenelse{\boolean{EpsNotation}}
  {(\wt J_\eps^1)^*} 
  {(\wt J^1)^*}   
} %

\title{First order operators on shrinking graph-like spaces}

\author{Pavel Exner}%
\address{Department of Theoretical Physics, NPI, Academy of Sciences,
  25068 \v{R}e\v{z} near Prague, and Doppler Institute, Czech
  Technical University, B\v{r}ehov\'{a}~7, 11519 Prague, Czechia}
\email{exner@ujf.cas.cz}

\author{Olaf Post}%
\address{Fachbereich 4 -- Mathematik, Universit\"at Trier, 54286
  Trier, Germany}%
\email{olaf.post@uni-trier.de}

\begin{document}

\begin{abstract}
  In this article we discuss the convergence of first order operators
  on a thickened graph (a graph-like space) towards a similar operator
  on the underlying metric graph.  On the graph-like space, the first
  order operator is of the form exterior derivative (the gradient) on
  functions and its adjoint (the negative divergence) on closed
  $1$-forms (irrotational vector fields).  Under the assumption that
  each cross section of the tubular edge neighbourhood is convex, that
  each vertex neighbourhood is simply connected and under suitable
  uniformity assumptions (which hold in particular, if the space is
  compact) we establish generalised norm resolvent convergence of the
  first order operator on the graph-like space towards the one on the
  metric graph.  The square of the first order operator is of Laplace
  type; on the metric graph, the function ($0$-form) component is the
  usual standard (Kirchhoff) Laplacian.  A key ingredient in the proof
  is a uniform Gaffney estimate: such an estimate follows from an
  equality relating here the divergence operator with all (weak)
  partial derivatives and a curvature term, together with a
  (localised) Sobolev trace estimate.
\end{abstract}

\maketitle

\setboolean{EpsNotation}{true}

%
\section{Introduction}
\label{sec:intro}
%

The problem of relations between the dynamics of a graph and on its
`fattened' version has a long history reaching back to the 1950s when
Ruedenberg and Scher~\cite{ruedenberg-scherr:53} provided a heuristic
justification of Pauling's model of aromatic hydrocarbon molecules
\cite{pauling:36}.  While the idea of a `fat graph' shrinking to its
skeleton was intuitively appealing, mathematically the problem
appeared to be challenging and it took time before rigorous versions
of the Ruedenberg and Scher argument were worked
out~\cite{kuchment-zeng:01, rubinstein-schatzman:01, exner-post:05};
namely the convergence of eigenvalues of the Neumann Laplacian towards
the standard Laplacian on the metric graph.  A sort of weak resolvent
convergence on weighted trees was shown in~\cite{saito:00}.  Later,
the second author developed a concept of generalised norm resolvent
convergence in~\cite{post:06} and applied it to the setting of
fattened graphs; these results were extended and generalised into
spectral analysis on graph-like spaces and into convergence of
operators acting in different Hilbert spaces~\cite{post:12}.  As a
consequence of the generalised norm resolvent convergence, one has
spectral convergence, see \App{que} for a brief overview.  Later, we
managed to treat graph dynamics going beyond the simplest (Kirchhoff)
vertex coupling (the standard Laplacian) in
graphs~\cite{exner-post:07,exner-post:09, exner-post:13}.

In contrast to the Schr\"odinger case, Dirac operators on graphs were
first considered as a purely mathematical
problem~\cite{bulla-trenkler:90}.  This changed with the advent of
graphene in which the electron dynamics, albeit non-relativistic, is
effectively described by the massless Dirac equation; for discussion
of physical effects like quantum chaos exhibited by Dirac particles on
metric graphs see, e.g.~\cite{bolte-harrison:03, harrison-winn:12} and
for the non-relativistic limit of such systems~\cite{bct:21}.

On the other hand, thin tubes supporting Dirac operators which can
model, for instance, graphene ribbons have been
studied~\cite{bbkob:22, bkob:23, exner-holzmann:22}, with the
particular attention paid to the situation when their perpendicular
size shrinks to zero; another limit investigated concerned Dirac
operator in a thin layer with infinite mass boundary conditions
shrinking to a smooth compact hypersurface in $\R^n$ without boundary
\cite{LOB25}.  Little is known, however, about the squeezing limit
when the tubes are a part of network with non-trivial branching
segments.  The problem of convergence (or divergence) of the
differential form \emph{spectrum} has been considered
in~\cite{egidi-post:17}, and this article can also be seen as a
continuation, showing the convergence of the corresponding
\emph{operators}.

\begin{figure}[h]
  \centering
  \begin{picture}(0,0)%
    \includegraphics{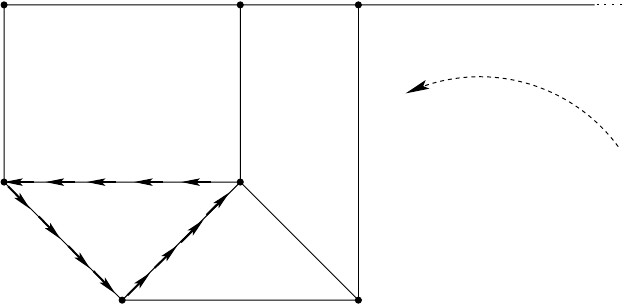}%
  \end{picture}%
  \setlength{\unitlength}{4144sp}%
  \begin{picture}(4768,2312)(420,-1892)
    \put(3196,-421){$\Xzero[e]$}%
    \put(4906,-826){$\interval0{\ell_e}$}%
    \put(3196,-1816){$v$}%
    \put(4186,164){$\Xzero[e']$}%
    \put(496,-1816){$\Xzero$}%
    \put(946,-871){$\Ff$}%
    \put(4321,-466){$\Psi_e$}%
  \end{picture}%
  \caption{A metric graph $\Xzero$ embedded in $\R^2$ with a
    coordinate map $\map{\Psi_e}{I_e}X$ (cf.\ \Def{mg}), an edge
    $\Xzero[e]$ of finite length $\ell_e \in \interval[open] 0\infty$
    and a semi-infinite edge $\Xzero[e']$.  The arrows indicate an
    harmonic vector field $\Ff$ ($\protect \mGradAdj \Ff=0$) on one loop.}
  \label{fig:metric-graph}
\end{figure}
The aim of the present note is to analyse such a limit when a network
collapses on its graph skeleton. We suppose that the structure
supports an abstract Dirac-type operator defined in terms of the
exterior derivative and its adjoint; in this framework we establish
norm resolvent convergence and estimate the convergence rate.
\begin{figure}[h]
  \centering
  \begin{picture}(0,0)%
    \includegraphics{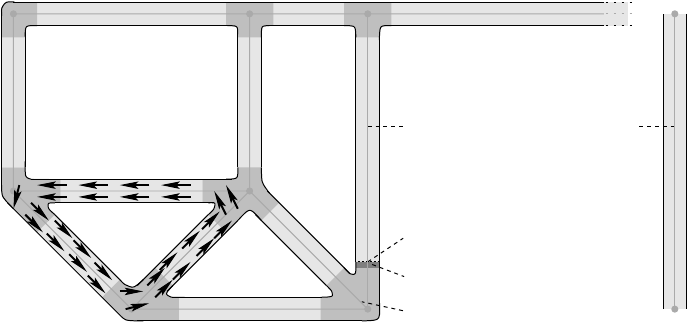}%
  \end{picture}%
  \setlength{\unitlength}{4144sp}%
  \begin{picture}(5245,2521)(348,-2029) \put(4636,119){$\Xepsed[']$}%
    \put(3421,-556){$\wtXepsed$}%
    \put(3421,-1951){$\Xepsvx$}%
    \put(406,-1906){$\wtXeps$}%
    \put(4816,-556){$\Xepsed$}%
  \end{picture}%
  \caption{\emph{Left:} An embedded associated graph-like space
    $\wtXeps$ with an (embedded) edge neighbourhood
    $\wtXepsed \cong \interval 0 {(1-\eps \tau)\ell_e}\times
    [-\eps/2,\eps/2]$ and vertex neighbourhood $\Xepsvx$ (cf.\
    \Def{gs} for the unscaled version describing the decomposition,
    \Def{gs.shrinking} for the shrinking parameter, \Def{gs.unif} for
    some uniformity assumptions and \Def{gs.emb} for the embedded
    case).  We also have drawn a harmonic $1$-form (vector field).
    \emph{Right:} the abstract vertex neighbourhood $\Xepsed$ of
    original length $\ell_e$.}
  \label{fig:graph-like-space}
\end{figure}
We define formally a metric graph $\Xzero$ in \Def{mg}.  If $\Xzero$
is embedded in $\R^m$ with straight edges (see \Fig{metric-graph}), we
may think of a shrinking family of graph-like spaces
$(\wtXeps)_{\eps \in \interval[open left] 0{\eps_0}}$ for some
$\eps_0 \in \interval[open left] 01$ as a (smoothened)
$\eps/2$-neighbourhood of $\Xzero$.  Here, ``smoothened'' means that
near the vertices we modify the boundary such that $\bd \wtXeps$ is of
class $\Contspace[2]$.  A graph-like space is basically such a set
where each vertex $v$ corresponds to a so-called \emph{vertex
  neighbourhood} $\Xepsvx$ and each edge $e \in E$ to an \emph{edge
  neighbourhood} $\wtXepsed$ (see \Def{gs} for the decomposition of
$\Xeps$ according to the graph and \Def{gs.shrinking} for the version
with the shrinking parameter $\eps$).  The decomposition is made in
such a way that a vertex neighbourhood is $\eps$-homothetic with a
fixed set $\Xonevx$, symbolically written as $\Xepsvx=\eps\Xonevx$.
The embedded case is easier explained, but technically more
complicated: for our first main result, we work with edge
neighbourhoods $\Xepsed$ of \emph{full} length $\ell_e$, i.e., with an
abstract graph-like space.  Moreover, as we do not assume that our
spaces are compact, we need some uniformity assumptions given in
\Def{gs.unif}; they are automatically fulfilled if the graph-like
space is compact.

Our first main result is as follows:
\begin{maintheorem}[generalised norm resolvent convergence of first
  order operators on abstract graph-like spaces]
  \label{thm:main-a}
  Let $\Xzero$ be a connected metric graph and
  $(\Xeps)_{\eps \in\interval[open left]01}$ an associated uniform
  family of graph-like spaces as in \Defss{gs}{gs.shrinking}{gs.unif}.
  Moreover, assume that the graph-like space is convex, i.e., that the
  boundary of the cross sections of the edge parts is convex and each
  vertex neighbourhood $\Xepsvx$ is simply connected
  (\Def{gs.convex}), then the first order operator $\xDirac$ on the
  graph-like space (restricted to closed $1$-forms, i.e., irrotational
  vector fields) converges in generalised norm resolvent sense to the
  first order operator $\mDirac$ on the metric graph.  The convergence
  rate is of order $\Err(\eps^{1/2})$.
\end{maintheorem}

Next, we consider a graph-like space $\wtXeps$ \emph{embedded} in
$\R^m$ as in \Fig{graph-like-space}: Let $\Xzero$ be a metric graph
embedded in $\R^m$ such that the edges correspond to straight line
segments.  We think of $\wtXeps$ as being a smoothened version of the
$\eps/2$-neighbourhood of $\Xzero$ (for details see \Def{gs.emb}).
Our second main result states that we can consider a ``real world''
example of an embedded graph-like space as perturbation of the
abstract graph-like space:
\begin{maintheorem}[embedded graph-like space is perturbation of
  abstract one]
  \label{thm:main-b}
  Let $\Xzero$ be a connected metric graph embedded in $\R^m$ and let
  $(\wtXeps)_{\eps \in\interval[open left]01}$ an associated uniform
  family of embedded convex graph-like spaces.  Then the corresponding
  first order operator $\wtxDirac$ is $\delta_\eps'$-quasi-unitarily
  equivalent with the first order operator $\xDirac$ on the abstract
  graph-like space as in the previous theorem with
  $\delta_\eps'=\Err(\eps^{1/2})$.
\end{maintheorem}
As a consequence of \Thms{main-a}{main-b} we obtain
\begin{maincorollary}[generalised norm resolvent convergence of first
  order operators on embedded spaces]
  \label{cor:main-c}
  Under the above assumptions, the first order operator $\wtxDirac$ on
  the embedded graph-like space converges in generalised norm
  resolvent sense to the first order operator $\mDirac$ on the
  embedded metric graph.
\end{maincorollary}

Another corollary of our first order convergence is the convergence of
the associated Laplacians on $0$- and $1$-forms
$\xDirac^2=\xLAPL=\xlapl\oplus\xLapl$ on the graph-like space towards
the Laplacian $\mDirac^2=\mLAPL=\mlapl\oplus\mLapl$ on the metric
graph.  Here, $\xlapl \ge 0$ is the Neumann Laplacian on $\Xeps$ while
$\mlapl \ge 0$ is the standard (``Kirchhoff'') Laplacian on the metric
graph $\Xzero$:
\begin{maincorollary}[generalised norm resolvent convergence of
  corresponding Laplacians]
  \label{cor:main-d}
  Under the above assumption the associated $p$-form Laplacians on the
  graph-like space (embedded or not) converge in generalised norm
  resolvent sense to the $p$-form Laplacian on the metric graph with
  convergence rate of order $\Err(\eps^{1/2})$ for $p=0$ or $p=1$.
\end{maincorollary}

The concept of $\delta$-quasi-unitary equivalence can be seen as
measuring a sort of ``distance'' between two operators (the resolvents
here) acting in different spaces. It is expressed with the aid of
so-called \emph{identification operators}
\begin{align*}
  \map{\IdOp}{\mHS=\mhs \oplus \mHs&=\Lsqr \Xzero \oplus \Lsqr \Xzero}
  {\xHS=\xhs \oplus \xHs=\Lsqr \Xeps \oplus \xHs}\\
  \FF = (\ff,\Ff) \mapsto
  \IdOp \FF &= (\idop \ff, \Idop \Ff),
\end{align*}
\begin{figure}[h]
  \centering
  \begin{picture}(0,0)%
    \includegraphics{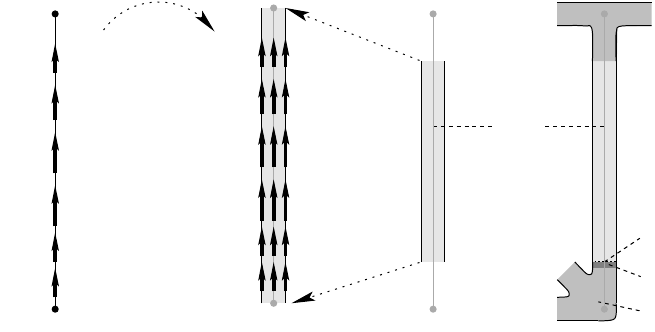}%
  \end{picture}%
  \setlength{\unitlength}{4144sp}%
  \begin{picture}(4977,2520)(211,-1804)
    \put(5131,-1726){$\Xepsvx$}%
    \put(5131,-1141){$\bd_v\wtXepsed=\bd_e\Xepsvx$}%
    \put(5131,-1411){$\Xepsed[,v]$}%
    \put(1801,-1726){$\Xepsed$}%
    \put(4051,-286){$\wtXepsed$}%
    \put(226,-1726){$\Xzero[e]$}%
    \put(1621,299){$\Idop\Ff$}%
    \put(811,299){$\Ff$}%
    \put(2836,479){$\Phi_e$}%
  \end{picture}%
  \caption{\emph{From left to right; first:}
    a metric edge $\Xzero[e]$
    with vector field $\Ff_e$.  The identification operator
    $\Idop$
    extends $\Ff_e$ constantly in the cross sectional
    direction. \emph{Second:} the edge neighbourhood of full length
    $\Xepsed$.  \emph{Third:} The shortened edge neighbourhood
    $\wtXepsed$ from the embedding together with the coordinate change
    map $\map{\Phi_e}{\wtXepsed}{\Xepsed}$ from \Subsec{rest.proofs}.
    \emph{Fourth:} a piece of the embedded graph-like space $\wtXeps$
    and the collar neighbourhood of the vertex neighbourhood
    $\Xepsvx[,e]$ of $\Xepsvx$ (dark grey) near the touching boundary
    $\bd_v \wtXepsed=\bd_e \Xepsvx$ (see
    \Defenum{gs.unif}{gs.unif.b}.}
  \label{fig:notation}
\end{figure}
where
$\xHs= \Sob[{\ExtDer[1]=0}] {T^*\Xeps} := \set{\Ww \in
  \Sob[{\ExtDer[1]}]{T^*\Xeps}} {\ExtDer[1] \Ww=0}$ is the closed
subspace of \emph{irrotational} vector fields (1-forms) on $\Xzero$.
Here, we basically extend functions and vector fields constantly in
the cross-sectional direction (see \Fig{notation}).  If
$\delta=\delta_\eps \to 0$ as $\eps \to 0$, one obtains a generalised
version of norm resolvent convergence.  As a consequence of
generalised norm resolvent convergence we mention convergence of the
spectrum and of suitable operator functions such as the heat operators
or spectral projections, see \App{que} for more details.

The assumption on the (edge) cross sectional spaces $\Yepsed$ to be
convex implies that $\Yepsed$ is simply connected; moreover for the
vertex neighbourhoods $\Xepsvx$ we assume that they are also simply
connected.  This assumption is very natural, as then, the limit metric
graph $\Xzero$ is a deformation retract of each graph-like space
$\Xeps$ for all $\eps>0$.  In particular, the homology groups
$H_p(\Xeps)$ of order $p \in \{0,1\}$ agree for all
$\eps \in \interval 01$, and in particular, the $p$-th Betti numbers
(dimension of $H_p(\Xeps)$) agree and are given by
\begin{equation}
  \label{eq:homology}
  \dim H_0(\Xeps)=1,
  \qquad
  \dim H_0(\Xeps)=\card E - \card V + 1.
\end{equation}
Moreover the indices of $\xDirac$ agree for all $\eps \in \interval01$, namely
\begin{equation}
  \label{eq:euler}
  \chi(\Xeps)=\dim H_0(\Xeps)-\dim H_1(\Xeps)
  = \card V-\card E.
\end{equation}
Note that we have restricted the forms on the graph-like space to
closed $1$-forms, hence no higher dimensional Betti numbers are
involved.

Whether elements in the domain of a first order operator like the
divergence fulfil additional regularity, for example being in
$\Sobspace$, is usually referred to as \emph{regularity}.  Even for
piecewise smooth domains, the divergence operator restricted to vector
fields in $\Sobspace$ might not be closed, in particular, if the
domain has ``re-entrant edges'' (cf.~\cite{birman-solomyak:87}).  A key
argument in our analysis is a formula by
Grisvard~\cite[Thm.~3.1.1.1]{grisvard:85} relating the divergence
operator with all weak derivatives and a curvature term
(see~\eqref{eq:gaffney.a}) for domains with $\Contspace[2]$-boundary.
A similar formula is shown in~\cite{gallot-meyer:88} for the form
Laplacian on general Riemannian manifolds with boundary.  Form
Laplacians on convex domains in Riemannian manifolds have also been
treated by Mitrea in~\cite{mitrea:01}; and more generally for
``almost'' convex Lipschitz domains in ~\cite{mtv:05}.  Recently,
Prokhorov and Filonov~\cite{prokhorov-filonov:15} proved
$\Sobspace$-regularity for ``weak'' convex domains in $\R^3$ (``weak''
here means that the domain is weakly
$\SobWspace[3]2 \cap \SobWspace[\infty] 1$-diffeomorphic to a convex
domain) Lamberti and Zaccaron~\cite{lamberti-zaccaron:23} proved
uniform Gaffney estimates in a different spirit: they control the
boundary being locally a graph of a function under uniform control.
Such estimates are used to get spectral stability of certain Maxwell
operators~\cite{lamberti-zaccaron:23}. In~\cite{lamberti-zaccaron:21}
they analysed the behaviour of such eigenvalues under domain
perturbation (see also the references therein).

It would be natural to consider also Lipschitz domains as for proper
$\eps$-neighbourhoods of metric graphs embedded for example in $\R^2$,
but for any vertex neighbourhood of a vertex of degree larger than
$2$, there would be an obtuse angle (``re-entrant edge'' according
to~\cite{birman-solomyak:87}, see also~\cite{birman:87b}), and for
such angles the Gaffney estimate no longer holds; the domain is only
a subset of $\Sobspace[1/2]$, see also Costabel and Dauge for
examples~\cite{costabel-dauge:00}.

\subsection*{Structure of the article}
In \Sec{notation} we describe the necessary notation (abstract Dirac
operators, metric graphs).  Moreover, we recall some facts on first
order operators on (flat) manifolds; including Kato's inequality
(\Lem{kato.ineq}) and the important Gaffney estimate~\Prp{gaffney} and
the key ingredient in our proof, a uniform Gaffney estimate
(\Thm{gaffney.scaled}).  In \Subsec{graph-like}, we then define the
various kinds of graph-like manifolds and recall some known estimates
on them.  \Sec{proofs} contains the proofs of our main results.  In
\App{que}, we briefly recall the notion of quasi-unitary equivalence
and generalised norm resolvent convergence; as well as a new result
suited to first order operators (\Prp{conv.lapl}).

\subsection*{Acknowledgements}
The research was supported by by the European Union’s Horizon 2020
research and innovation programme under the Marie Sk\l odowska-Curie
grant agreement No 873071.  We would like to thank Gilles Carron for
pointing our attention to~\cite{gallot-meyer:88} for the geometric
meaning of elliptic estimates and Pier Domenico Lamberti and Michele
Zaccaron for providing us with further literature and a helpful
discussion concerning uniform Gaffney estimates.  Last but not least
we also thank the referees for carefully reading and useful
suggestions.
%
\section{First order operators on metric graphs and graph-like spaces}
\label{sec:notation}
%


\subsection{First order operators abstractly}
\label{ssec:1st.order}

We now recall the notion of exterior derivative in an abstract Hilbert
space setting following~\cite[Sec.~1.2]{post:09c}
\begin{definition}[abstract exterior derivative and associated first
  order operator]
  \label{def:ext.der}
  Assume that
  \begin{equation}
    \label{eq:hs.1st}
    \HS
    = \hs \oplus \Hs
  \end{equation}
  is a Hilbert space.  Here, $\FF=(\ff,\Ff) \in \HS$ consists of the
  $0$-form (function) component $\ff \in \hs$ resp.\ the $1$-form
  (vector field) component $\Ff \in \Hs$.

  Let $\hsdom \subset \hs$.  An operator $\map \Grad {\hsdom} {\Hs}$
  is called an \emph{exterior derivative} if $\hsdom$ is dense in
  $\hs$ and if it is closed as operator $\hs \to \Hs$, i.e.,if
  $\hsdom$ with norm given by
  $\normsqr[\hsdom] {\ff} := \normsqr[\Hs]{\Grad
    \ff}+\normsqr[\hs]{\ff}$ is a Hilbert space.

  The associated \emph{first order operator} is defined by
  \begin{equation}
    \label{eq:first.order}
    \HSdom
    = \hsdom \oplus \Hsdom,
    \qquad
    \map{\Dirac}{\HSdom}{\HS}
  \end{equation}
  with action in matrix notation (with respect to the decomposition
  $\HS=\hs \oplus \Hs$)
  \begin{equation}
    \label{eq:first.order'}
    \Dirac=
    \begin{pmatrix}
      0 & \GradAdj\\
      \Grad & 0
    \end{pmatrix},
    \qquad
    \Dirac
    \FF =
    \begin{pmatrix}
      0 & \GradAdj\\
      \Grad & 0
    \end{pmatrix}
    \begin{pmatrix}
      \ff\\ \Ff
    \end{pmatrix}
    =
    \begin{pmatrix}
      \GradAdj \Ff\\
      \Grad \ff
    \end{pmatrix}.
  \end{equation}
  Here, $\Hsdom = \dom \GradAdj \subset \Hs$ with norm given by
  $\normsqr[\Hsdom] {\Ff} := \normsqr[\hs]{\GradAdj
    \Ff}+\normsqr[\Hs]{\Ff}$ is again a Hilbert space.
\end{definition}
It is easily seen that $\Dirac$ is self-adjoint and that $\HSdom$
becomes a Hilbert space with norm defined by
$\normsqr[\HSdom] \FF := \normsqr[\HS] {\Dirac \FF} + \normsqr[\HS]
\FF$.  Note that we have
\begin{subequations}
  \label{eq:norm.dirac}
  \begin{align}
    \nonumber
    \normsqr[\HSdom] \FF
    &=\normsqr[\HS] {\Dirac \FF} + \normsqr[\HS] \FF\\
    \label{eq:norm.dirac.a}
    &=\iprod[\HS]{(\Dirac^2+1)\FF}{\FF}
      = \iprod[\HS]{(\Dirac \pm \im)(\Dirac \mp \im)\FF}{\FF}
      =\normsqr[\HS]{(\Dirac \mp \im)\FF} \quad\text{and}\\
    \label{eq:norm.dirac.b}
    \normsqr[\HSdom] \FF
    &=\normsqr[\hs]{\ff} + \normsqr[\Hs]{\Ff} +
      \normsqr[\hs]{\GradAdj \Ff} +\normsqr[\Hs]{\Grad \ff}
      =\normsqr[\hsdom]{\ff} +\normsqr[\Hsdom]{\Ff}
  \end{align}
\end{subequations}
first for all $F \in \dom (\Dirac^2)$ and then by density for
$F \in \HSdom=\dom \Dirac$.  We define the \emph{Laplacian} associated
with $\Dirac$ by $\LAPL := \Dirac^2$.

Also $\LAPL$ is self-adjoint and acts as
\begin{equation}
  \label{eq:lapl}
  \LAPL =
  \begin{pmatrix}
    \lapl & 0\\
    0 & \Lapl
  \end{pmatrix},
  \qquadtext{where}
  \lapl=\GradAdj\Grad
  \quadtext{and}
  \Lapl=\Grad \GradAdj.
\end{equation}
Moreover, $\lapl$ and $\Lapl$ have the same spectrum (except in $0$)
including multiplicity (see e.g.~\cite[Prop.~1.2]{post:09c}).
Recalling the analogy with the usual Dirac operator, one can think of
$\Dirac$ as of an \emph{abstract Dirac-type operator}.  Moreover,
$\ker \lapl =\ker \Grad$ and $\ker \Lapl=\ker \GradAdj$.

We have the abstract Hodge decomposition
\begin{equation}
  \label{eq:hodge}
  \HS = \ker \Dirac \oplus \clo{\ran \GradAdj} \oplus \clo{\ran \Grad}, \qquad
  \hs = \ker \Grad \oplus \clo{\ran \GradAdj}
  \quadtext{and}
  \Hs = \ker \GradAdj \oplus \clo{\ran \Grad}
\end{equation}
If $0$ is isolated in $\spec \Dirac$ then the ranges $\ran \Grad$ and
$\ran \GradAdj$ are closed in $\Hs$ and $\hs$ and we can omit the
closures for the ranges $\ran \GradAdj$ and $\ran \Grad$.  An
equivalent condition for $0$ to be isolated in $\spec \Dirac$ is that
$0$ is isolated in $\spec \lapl$ or $\spec \Lapl$, or in other words
\begin{equation}
  \label{eq:spec.cond}
  \inf (\spec \lapl \setminus\{0\})>0
  \quadtext{or}
  \inf (\spec \Lapl \setminus\{0\})>0
\end{equation}
(see \cite[Prop.~1.3]{post:09c}).  The \emph{index} of a Dirac
operator $\Dirac$ is defined once $\dim \ker \Grad$ and
$\dim \ker \GradAdj$ are finite, and is given by
\begin{equation}
  \label{eq:index}
  \ind \Dirac
  := \dim \ker \Grad - \dim \ker \GradAdj.
\end{equation}

\subsection{Metric graphs, related Hilbert spaces and operators}
\label{ssec:met.graphs}

As a simple example of how to implement the abstract theory of the
preceding subsection we start with first order operators on a metric
graph.
\begin{definition}[metric graph]
  \label{def:mg}
  Let $\Xzero$ be a connected metric space and $E$ an at most countable
  set.  We say that $\Xzero$ is a \emph{metric graph} if
  \begin{enumerate}
  \item there is a map
    $\map \ell E {\interval{\ell_0}\infty}$ for some
    $\ell_0>0$, called \emph{length function}; without loss of
    generality, we assume that $\ell_0=\inf_{e \in E} \ell_e$;
  \item for each $e \in E$ there is an isometry
    $\map{\Psi_e}{I_e}\Xzero$ from $I_e:=\interval 0{\ell_e}$ (or
    $I_e:=\interval[open right] 0 \infty$ if $\ell_e=\infty$) onto
    $\Xzero[e]:=\Psi_e(I_e)$, called \emph{(edge) coordinate};
    $\Xzero[e]$ is called \emph{(metric) edge}; the edge coordinate
    introduces an \emph{orientation} on $\Xzero[e]$;
  \item if $\ell_e<\infty$ we assume that $\card{\bd \Xzero[e]}=2$
    (there are no loop-shaped edges); edges with $\ell_e=\infty$ are
    called \emph{semi-infinite} edge;
  \item the coordinates cover $\Xzero$, i.e.,
    $\Xzero=\bigcup_{e \in E} \Xzero[e]$; and
    $\intrXzero[e] \cap \intrXzero[e'] =\emptyset$ whenever $e \ne e'$;
  \item $V:=\bigcup_{e \in E} \bd \Xzero[e]$ is discrete in $\Xzero$,
    called the \emph{vertex set};%
  \item $E_v := \set{e \in E}{v \in \Xzero[e]}$ is the set of edges
    \emph{adjacent} with the vertex $v$; and $\deg v := \card{E_v}$ is
    called the \emph{degree} of $v$; we assume here that $\deg v$ is
    finite (i.e., the graph is locally finite);
  \item $\Xzero$ is endowed with a measure $\lebesgue$ such that
    $\lebesgue(B)=\sum_{e \in E} \lebesgue_e(\Psi_e^{-1}(B))$ for all Borel
    subsets $B \subset \Xzero$; here $\lebesgue_e$ denotes the Lebesgue
    measure on $I_e \subset \R$.
  \end{enumerate}
\end{definition}
We need a positive uniform lower bound on the edge lengths (in case the
graph is infinite):
\begin{definition}[uniform metric graph]
  \label{def:mg.unif}
  We say that a metric graph $\Xzero$ has \emph{uniformly separated
    vertices} if
  \begin{equation}
    \label{eq:ell0}
    \ell_0 := \inf_{e \in E} \ell_e>0.
  \end{equation}
  For the sake of brevity we will call such graphs simply \emph{uniform}.
\end{definition}

A function $\map f \Xzero \C$ is entirely determined by its values
$f_e := f \circ \Psi_e$ ($e \in E$) and we often simply identify $f$
with the family $(f_e)_{e \in E}$.

According to the edge coordinates, we have a natural (unitary)
identification of the Hilbert space
\begin{equation}
  \label{eq:mg.dec}
  \Lsqr{\Xzero,\lebesgue}
  \cong \bigoplus_{e \in E} \Lsqr{I_e,\lebesgue_e}
\end{equation}
via $f \mapsto (f_e)_{e \in E}$.  In the sequel, we omit the measure
in the $\Lsqrspace$-spaces, i.e.,we simply write $\Lsqr \Xzero$
instead of $\Lsqr{\Xzero,\lebesgue}$ or $\Lsqr{I_e}$ instead of
$\Lsqr{I_e,\lebesgue_e}$.

We now consider the space of square-integrable \emph{forms} on a metric
graph given by
\begin{equation}
  \label{eq:lsqr.mg.forms}
  \mHS
  = \mhs \oplus \mHs
  \quadtext{where}
  \mhs=\Lsqr {\Xzero}
  \quadtext{resp.}
  \mHs=\Lsqr {\Xzero}.
\end{equation}
Here, $\FF=(\ff,\Ff) \in \mHS$ consists of the $0$-form (function)
component $\ff \in \mhs$ resp.\ $1$-form (vector field) component
$\Ff \in \mHs$.  Note that $0$- and $1$-forms on a metric graph are
formally the same due to the one-dimensional character of the edges,
they differ in their interpretation (and also in their vertex
evaluation).

In order to make a difference in notation we add $\dd s$ to a $0$-form
to turn it into a $1$-form, i.e., we define
\begin{equation}
  \label{eq:fct.vct}
  \map{(\cdot) \dd s}{\mhs}{\mHs},
  \qquad
  f=(f_e)_e \mapsto f \dd s=(f_e \dd s_e)_e
\end{equation}
with inverse given by
$\Ff \mapsto \Ff \cdot \dd s=(\Ff_e \cdot \dd s_e)_e$.  In
particular, we have $\Ff=(\Ff \cdot \dd s) \dd s$.  We also
define the following Sobolev spaces
$\Sobdec {\Xzero}:= \bigoplus_{e \in E} \Sob{I_e}$ of (decoupled)
square-integrable functions and first weak derivatives with norm
defined by
\begin{equation}
  \label{eq:norm.sob}
  \normsqr[\Sobdec{\Xzero}] f
  := \sum_{e \in E} \normsqr[\Sob{I_e}]{f_e}
  = \sum_{e \in E} \Bigl(\int_{I_e} \abssqr{f_e'(s_e)} \dd s_e
                          + \int_{I_e} \abssqr{f_e(s_e)} \dd s_e \Bigr),
\end{equation}
where $\dd s_e$ denotes integration with respect to the Lebesgue
measure $\lebesgue$ (we often omit the subscript $(\cdot)_e$ and
simply write $s$ or $\dd s$).  A function in $\Sob{I_e}$ is
continuous, hence it makes sense to speak about function values.
Moreover, we define
\begin{subequations}
  \begin{align}
    \label{eq:sob.mg.forms.a}
    \Sob {\Xzero}
    &= \bigset{(\ff_e)_e \in \Sobdec{\Xzero}}{\ff \in \Cont{\Xzero}}
      \quad\text{and}\\
    \label{eq:sob.mg.forms.b}
    \Sobsum {\Xzero}
    &= \bigset{(\Ff_e)_e \in \Sobdec{\Xzero}}
      {\sum_{e \in E_v} \Ff_e(v)=0}.
  \end{align}
\end{subequations}
Here, for functions, each component $\ff_e$ has the same value
$\ff_e(v)$ at a vertex $v \in \Xzero[e]$ for all $e \in E_v$ we denote it
$\ff(v)$.  For $1$-forms, we define the \emph{oriented evaluation} by
\begin{equation}
  \label{eq:or.eval}
  \Ff_e(v)=
  \begin{cases}
    \Ff_e(\ell_e) & \text{if $v=\Psi_e(\ell_e)$,}\\
    -\Ff_e(0) & \text{if $v=\Psi_e(0)$,}
  \end{cases}
\end{equation}
hence $1$-forms see the orientation of a graph when their values at a
vertex are evaluated; it ``flows'' out of the edge at a terminal
vertex ($v=\Psi_e(\ell_e)$) and ``flows'' into the edge at an initial
vertex ($v=\Psi_e(0)$).\footnote{We use a different sign convention
  for forms (or for derivatives $f_e'(v)$, see later) than
  in~\cite{berkolaiko-kuchment:13} or~\cite{kuchment:04}.  One reason
  is that with this convention, an integration by parts formula holds
  with the same signs as on intervals or manifolds, see
  e.g.~\cite[Eq.~(2.11)]{post:12}; also we prefer to have the Robin
  term on metric graphs to have the ``correct'' ($+$)-sign,
  cf.~\cite[Thm.~9]{kuchment:04}.}

We now define the \emph{exterior derivative} on a metric graph by
\begin{equation}
  \label{eq:ext.der.mg}
  \map{\mGrad}{\mhsdom=\Sob{\Xzero}}{\mHs=\Lsqr {\Xzero}},
  \qquad
  \ff \mapsto \mGrad \ff =(\ff_e' \dd s_e)_{e \in E}.
\end{equation}
\begin{lemma}
  \label{lem:ext.der}
  The operator $\mGrad$ with domain
  $\dom \mGrad = \mhsdom=\Sob{\Xzero}$ is closed as operator in
  $\mhs \to \mHs$.  In particular, it is an exterior derivative in the
  sense of \Def{ext.der}.  Its adjoint is given by
  \begin{equation}
    \label{eq:div}
    \map{\mGradAdj}{\Sobsum \Xzero}{\mhs},
    \qquad
    \Ff \mapsto \mGradAdj \Ff=(-\Ff_e'\cdot \dd s_e)_{e \in E}
  \end{equation}
  with domain $\dom \mGradAdj = \mHsdom=\Sobsum{\Xzero}$ given
  in~\eqref{eq:sob.mg.forms.b}.  The index of the associated
  Dirac-type operator $\mDirac$ is the Euler characteristic, i.e.,
  $\ind \mDirac=\card V - \card E$.
\end{lemma}
\begin{proof}
  The spaces $\Sob {\Xzero}$ and $\Sobsum {\Xzero}$ are both closed
  subspaces of $\Sobdec {\Xzero}$ as the evaluation maps
  \begin{align*}
    \ff \mapsto \ff(v) \qquadtext{and}
    \Ff \mapsto \Ff_e(v)
  \end{align*}
  are bounded from $\Sobsum {\Xzero}$ with norm bounded by
  $\cosh(\ell_0/2)^{1/2}$ (see e.g.~\cite[Sec.~6.1]{post:16}).  The
  statement about the adjoint follows by a straightforward
  calculation.  The index of $\mDirac$ was calculated e.g.\
  in~\cite[Sec.~6]{post:09c}.
\end{proof}

Note that the corresponding Laplace operator on $0$-forms is the usual
standard (also called Kirchhoff) Laplacian on a metric graph.


\subsection{First order operators on Euclidean spaces with boundary}
\label{ssec:1st.order.mfds}
Before fixing the corresponding setting of first order operators on
graph-like spaces, we briefly review some general facts about
$1$-forms and first order operators.  Let $X$ be a flat Riemannian
manifold of dimension $m \ge 2$ with boundary $\bd X$ of class
$\Contspace[2]$ (or $\Contspace[1,1]$); an example is a closed subset
of $\R^m$ with smooth boundary.  We use the language of manifolds here
and employ the symbol $g$ for the flat metric on $X$ (i.e.,$g$ is
locally Euclidean) for fixing some scaling arguments later.  The main
reason is that restricting to subsets of $\R^m$ would require some
technical details for graph-like spaces arising as neighbourhoods of
metric graphs $\Xzero$ embedded in $\R^m$, hence we formulate this
case as a perturbation of the abstract one, see \Thm{main-b}.

We write $\intr X$ for the interior of $X$, hence
$X=\intr X \dcup \bd X$.  Moreover, for a subset $X'$ of $X$, we
define the \emph{internal} boundary (i.e., the boundary in the
topology of $X$) by
\begin{equation}
  \label{eq:def.int.bd}
  \intbd X' := \clo {X'}  \cap \clo{X \setminus X'}
\end{equation}
where $\clo {(\cdot)}$ is the closure in $X$.

For integration with respect to the induced measure (Lebesgue measure
resp.\ hypersurface measure) we write $\dd X$ resp.\ $\dd \bd X$.  The
space of square-integrable (classes of) functions is denoted by
$\Lsqr X$.  Moreover, let $TX:=X \times (\C^n)^*$ the (complexified)
cotangent bundle; a $1$-form is now a section $\map \Uu X {T^*X}$

For some aspects of abstract boundary value theory with examples on
manifolds, see also~\cite{post:07}.

We write $\dd X$ resp.\ $\dd \bd X$ for the measure with respect to
the corresponding natural measure on $X$ resp.\ $\bd X$.  Denote by
\begin{equation}
  \label{eq:ext.der.mfd}
  \map \Grad {\Sob X}{\Lsqr{T^* X}}
\end{equation}
the exterior derivative, where $\Sob X$ is the space of weakly
differentiable functions with derivative in $\Lsqrspace$ with norm (of
Sobolev-type) defined by
$\normsqr[\Sob X] \uu := \normsqr[\Lsqr{T^*X}]{\Grad \uu} + \normsqr[\Lsqr
X] \uu$.  The formal adjoint of $\Grad$ is the (negative) divergence
operator
\begin{equation}
  \label{eq:div.mfd}
  \map \Div {\Sob[\Div] {T^*X}}{\Lsqr X},
  \quadtext{where}
  \Sob[\Div]{T^*X}:=
  \set{\Ww \in \Lsqr{T^*X}}{\Div \Ww \in \Lsqr X}.
\end{equation}
Here, $w=\Div \Ww \in \Lsqr X$ if
\begin{align*}
  \int_X w \, \conj \phi \dd X = \int_X \iprod[T^*X] \Ww {\Grad \phi} \dd X
\end{align*}
holds for all smooth functions $\phi \in \Cci X$ with compact support
in $\intr X$.  From the divergence theorem (for weakly differentiable
functions) and
\begin{align*}
  \Div(\uu \Ww)=\uu (\Div \Ww) - \iprod[T^*X] {\Grad \uu} \Ww
\end{align*}
we conclude the following integration by parts formula
\begin{equation}
  \label{eq:part.int}
  \int_X \iprod[T^*X]{\Grad \uu}\Ww \dd X
  = \int_X \uu (\Div \conj \Ww) \dd X
  + \int_{\bd X} \uu (\conj \Ww \cdot \normvec {\bd X}) \dd \bd X
\end{equation}
for all $\uu \in \Sob X$ and $\Ww \in \Sob[\Div] {T^*X}$ in the
distributional sense, where $\normvec {\bd X}$ denotes the outwards
normal vector on $\bd X$ and $\Ww \cdot \normvec{\bd X}$ the normal
component of $\Ww$ on $\bd X$.  Here,
$\Ww(\normvec{\bd X}) \in \Sob[-1/2]{\bd X}$.  Now, from
integration by parts one can see that
\begin{equation}
  \label{eq:dom.ext.der.adj}
  \dom \GradAdj
  = \set{\Ww \in \Sob[\Div]{T^*X}}{\Ww (\normvec{\bd X})=0},
\end{equation}
i.e., the $1$-form has only tangential components on $\bd X$.

\subsubsection*{Restriction to irrotational vector fields}
As on a metric graph there is no counterpart of a $2$-form, it turns
out that we have to restrict ourselves to $1$-forms with vanishing
exterior derivative.  In the language of three-dimensional vector
fields, the exterior derivative is the curl operator and such vector
fields are called \emph{irrotational}.

Denote by $\map{\ExtDer \Ww=\ExtDer[1] \Ww} X {\Forms 2 X}$ the
exterior derivative of the $1$-form $\Ww$.  We assume that
$\ExtDer[1] \Ww \in \Lsqr{\Forms 2 X}$, i.e., there is
$\beta=\ExtDer[1] \Ww \in \Lsqr{\Forms 2 X}$ such that
\begin{align*}
  \int_X \iprod[\Forms 2 X] \beta \eta \dd X
  = \int_X \iprod[T^*X] \Ww {\Div[1] \eta} \dd X
\end{align*}
holds for all smooth $2$-forms $\eta \in \Cci {\Forms 2 X}$ with
compact support in $\intr X$, where $\Div[1]$ denotes the formal
adjoint of $\ExtDer[1]$.  For $m \in \{2,3\}$ and $X \subset \R^m$,
the results are stated in~\cite[Ch.~I]{girault-raviart:86} but the
generalisation to arbitrary $m \in \N$ is straightforward (note that
$\dim \Forms[x] 2 X=\binom m 2=m(m-1)/2$).  Denote by
\begin{equation}
  \label{eq:rot.sob}
  \Sob[{\ExtDer[1]}]{T^*X}
  =\set{\Ww \in \Lsqr{T^*X}}
  {\ExtDer[1] \Ww \in \Lsqr{\Forms 2 X}}
\end{equation}
the largest space on which $\ExtDer[1]$ is defined in the
$\Lsqrspace$-sense.  We now define the Hilbert space of $1$-forms on
which we perform our abstract analysis as in \Subsec{1st.order}:  We set
\begin{equation}
  \label{eq:1-form.mfd}
  \Hs
  := \Sob[{\ExtDer[1]=0}] {T^*X}
  := \set{\Ww \in \Sob[{\ExtDer[1]}]{T^*X}}
  {\ExtDer[1] \Ww=0}.
\end{equation}
It can be seen that $\Hs$ is indeed a closed subspace of
$\Lsqr{T^*X}$.  Its orthogonal complement consists of ``fluxless
knots'' (in the terminology of~\cite{cdtg:02}); they belong to
$\ker \Div$ as they are ``divergence free''.

\subsubsection*{Regularity of first order operator domains}
We now turn to some delicate facts about the space
\begin{align*}
  \Sob[{\Div,\ExtDer[1]}]{T^*X}
  :=& \Sob[\Div]{T^*X} \cap \Sob[{\ExtDer[1]}]{T^*X}\\
  =& \bigset{\Ww \in \Lsqr{T^*X}}
  {\Div \Ww \in \Lsqr X, \; \ExtDer[1] \Ww \in \Lsqr{\Forms 2 X}}
\end{align*}
and its restriction
\begin{align*}
  \Sob[{\Div,\ExtDer[1]=0}]{T^*X}
  :=& \Sob[\Div]{T^*X} \cap \Sob[{\ExtDer[1]=0}] {T^*X}\\
  =& \bigset{\Ww \in \Lsqr{T^*X}}
     {\Div \Ww \in \Lsqr X, \; \ExtDer[1] \Ww =0}
     \subset \HS^1.
\end{align*}
What is now vital for our analysis is that if $\bd X \ne \emptyset$
then none of the above spaces is a subset of $\Sob{T^*X}$, where
$\Sob{T^*X}$ consists of $\Ww \in \Lsqr{T^*X}$ such that its covariant
derivative $\nabla \Ww$ is in $\Lsqr{\Tensor 2 X}$ (for the sake of
brevity, $\Tensor 2 X=T^* X \otimes T^* X$) with norm given by
\begin{equation}
  \label{eq:sob.vec.norm}
  \normsqr[\Sob {T^*X}] \Ww
  := \normsqr[\Lsqr{\Tensor 2 X}]{\nabla \Ww}
  + \normsqr[\Lsqr{T^*X}]\Ww
\end{equation}
If $X \subset \R^m$, then $\normsqr[\Sob {T^*X}] \Ww$ is just the
component-wise $\Sobspace$-norm of $\Ww$.

\begin{lemma}
  \label{lem:triv.emb}
  We have
  \begin{align*}
    \normsqr[\Lsqr X]{\Div \Ww}
    \le m \sum_{j=1} \normsqr[\Lsqr X]{\partial_j \Ww_j}
    \le m \normsqr[\Lsqr{\Tensor 2 X}] {\nabla \Ww}
  \end{align*}
  where $m=\dim X$ and $\Ww=\sum_{j=1}^m \Ww_j \dd x_j$.  In
  particular, $\Sob[\Div]{T^*X} \subset \Sob{T^*X}$.
\end{lemma}
\begin{proof}
  Fix an orthonormal frame in $X$ (as $X$ is globally flat, such an
  orthonormal frame exists globally).  Let $(\Ww_j)_{j=1,\dots,n}$ be
  the coordinates of $\Ww$ with respect to this orthonormal frame.
  Then we have $\Div \Ww=-\sum_{j=1}^m \partial_j \Ww _j$ hence
  \begin{align*}
    \bigabssqr{\Div \Ww}
    = \Bigabssqr{\sum_{j=1}^m (-1) \cdot \partial_j \Ww_j}
    \le m \sum_{j=1}^m \abssqr{\partial_j \Ww_j}
  \end{align*}
  using Cauchy-Schwarz and the result follows by integrating over $X$.
\end{proof}

\subsubsection*{Kato's inequality}
We state a useful estimate (and give its simple proof) in order to
``recycle'' our scalar estimates already done on graph-like spaces
(see e.g.~\cite[Ch.~6]{post:12}); it is sometimes also called Kato's
inequality, it was already used in~\cite[Prop.~3.9]{anne-post:21}:
\begin{lemma}[Kato's inequality]
  \label{lem:kato.ineq}
  Let $E$ be a complex vector bundle over $X$ with a (sesquilinear
  bundle) metric $h$ and a metric covariant derivative $\nabla$.  Let
  $\omega \in \Sob{T^*X}$ then $\abs[h] \omega \in \Sob X$ and
  \begin{equation}
    \label{eq:kato.ineq}
    \bigabs[g]{d \abs[h] \omega}
    \le \abs[h]{\nabla\omega},
  \end{equation}
  where
  $\map{\abs[h]\omega=\sqrt{h(\omega,\omega)}} X {\interval[open
    right] 0 \infty}$ and where $\abs[g]\cdot$ is the induced norm on
  $T^*X$ from the metric on $X$.
\end{lemma}
\begin{proof}
  Let $\xi \in T_xX$ be a tangent vector.  As $\nabla$ is a metric
  covariant derivative, we have
  \begin{align*}
    d(h(\omega,\omega)(\xi)
    =h(\nabla_\xi\omega,\omega)+h(\omega,\nabla_\xi\omega)
    =2\Re h(\nabla_\xi\omega,\omega),
  \end{align*}
  hence we have
  \begin{align*}
    d \abs[h]\omega(\xi)
    =d \sqrt{h(\omega,\omega)}
    =\frac1{\sqrt{h(\omega,\omega)}} \Re h(\nabla_\xi\omega,\omega)
    \le \abs[h]{\nabla_\xi\omega}
  \end{align*}
  using Cauchy-Schwarz for the inequality.  Taking now the pointwise
  operator norm in $T^*X$ (denoted by $\abs[g]\cdot$) we obtain the
  desired inequality.
\end{proof}

The Kato inequality can be used as follows:
\begin{corollary}
  \label{cor:kato.ineq}
  Let $E$, $h$ and $\nabla$ be as before.  Suppose that
  $X_1 \subset X_2 \subset X$ and that there are $C,C'\ge 0$ such that
  \begin{align*}
    \normsqr[\Lsqr {X_1,g}] u
    \le C' \normsqr[\Lsqr{T^*X_2,g}]{d u}
    + C \normsqr[\Lsqr{X_2,g}]u
  \end{align*}
  holds for all $u \in \Sob {X_2}$, then we have
  \begin{align*}
    \normsqr[\Lsqr{E \restr {X_1},h}]\omega
    \le C' \normsqr[\Lsqr{(T^*X \otimes E)\restr{X_2},g \otimes h}]{\nabla\omega}
    + C \normsqr[\Lsqr{E \restr{X_2},g}]\omega
  \end{align*}
  for all $\omega \in \Sob{E,h}$.
\end{corollary}

\subsection{Regularity and Gaffney estimates}
\label{ssec:gaffney}

\subsubsection*{Second fundamental form}
For the Gaffney estimates, we need to define the second fundamental
form $\II_{\bd X}$ of $\bd X$ in $X$: recall that
$\map {\normvec{\bd X}}{\bd X} {TX}$ is the normal vector field, and
that $T_xX=T_x\bd X\oplus \C \normvec {\bd X}(x)$.  We define now the
second fundamental form $\map {\II_{\bd X}}{T\bd X \times T\bd X} \R$
given by\footnote{We use the sign convention that characterises
  convexity of $\bd X$ in $X$ by $\II_{\bd X} \ge 0$.}
\begin{equation}
  \label{eq:2ff}
  x \mapsto \II_{\bd X,x}(\xi,\eta)
  =g_x(\nabla_\xi\normvec {\bd X,x},\eta)
\end{equation}
for $\xi, \eta \in T_x\bd X$, where $\nabla$ is the (Levi-Civita)
covariant derivative on $TX$.  It can be seen that
$\nabla_\xi \normvec \bd X \in T\bd X$; actually, $\II_{\bd X}$ is
symmetric, hence it is real-valued.  In the next definition, we extend
the second fundamental form to $1$-forms on $\bd X$.

We also define the integral over the fundamental form on the boundary as
\begin{equation}
  \label{eq:2ff.int}
  S_{\bd X}(\Uu,\Ww)
  := \int_{\bd X} \II_{\bd X}(\iota^*\Uu,\iota^*\Ww) \dd \bd X
\end{equation}
for $\Uu, \Ww \in \Sob{T^*X}$, where $\embmap \iota {\bd X} X$ is the
canonical embedding.

We need a statement on the fundamental form on spaces which are
(locally) a product:
\begin{lemma}
  \label{lem:2ff.prod}
  Assume that $X$ in a neighbourhood of $x \in \bd X$ is isometric to
  a neighbourhood of $(s,y)$ in $\interval[open] 0 {\ell_0} \times Y$,
  where $y \in \bd Y$, and where $Y$ is a compact convex manifold,
  i.e., we have $\II_{\bd Y}\ge 0$ on $\bd Y$ then
  $\II_{\bd X,x} \ge 0$.
\end{lemma}
\begin{proof}
  The claim follows from a straightforward calculation using
  definition~\eqref{eq:2ff}.
\end{proof}
\subsubsection*{Gaffney estimates}
We now state the following result following
from~\cite[Thm.~3.1.1.1]{grisvard:85} adapted to our needs:
\begin{proposition}
  \label{prp:gaffney}
  \begin{subequations}
    Denote by $\map \Grad {\Sob X}{\Hs}$ the exterior derivative on
    functions into the closed forms (irrotational vector fields,
    cf.~\eqref{eq:1-form.mfd}), then we have
  \begin{enumerate}
  \item
    \label{gaffney.a}
    for $\Ww \in \Sob{T ^*X}$ with $\ExtDer[1] \ww=0$ and
    $\Ww (\normvec {\bd X})=0$ on $\bd X$ we have
    \begin{equation}
      \label{eq:gaffney.a}
      \normsqr[\Lsqr X]{\GradAdj \Ww}
      = \normsqr[\Lsqr{\Tensor 2 X}] {\nabla \Ww}
      + S_{\bd X}(\Ww,\Ww),
    \end{equation}
    where $S_{\bd X}$ is the integral over the second fundamental form
    of $\bd X$ in $X$, cf.~\eqref{eq:2ff.int}.
  \item
    \label{gaffney.b}
    \begin{equation}
      \label{eq:gaffney.b}
      \Hsdom:=
      \dom \GradAdj
      =\bigset{\Ww \in \Lsqr{T^*X}}
      {\Div \Ww \in \Lsqr X, \; \ExtDer[1] \Ww =0, \;
      \Ww(\normvec {\bd X})=0}
    \end{equation}
    is a subset of $\Sob{T^*X}$.  Moreover, the so-called \emph{Gaffney
      estimate} holds, i.e.
    \begin{equation}
      \label{eq:gaffney}
      \normsqr[\Lsqr{\Tensor 2 X}] {\nabla \Ww}
      \le \CGaffney \normsqr[\Hsdom] \Ww
    \end{equation}
    holds for all $\Ww \in \Hsdom$, where
    \begin{equation}
      \label{eq:gaffney.c}
      \CGaffney
      =
      \begin{cases}
        1, & \text{if $\bd X$ is convex in $X$, i.e., $\II_{\bd X} \ge 0$,}\\
        \max \{2,8(\kappa^-_{\bd X})^2\}, & \text{otherwise,}
      \end{cases}
    \end{equation}
    and where $\kappa^-_{\bd X}$ 
    is the supremum of the absolute value of the minimal negative
    principal curvature of $\bd
    X$, 
    i.e., we have
    \begin{equation}
      \label{eq:2ff.lower}
      \kappa^-_{\bd X}
      := \sup_{x \in \bd X}\sup_{\xi \in T_x^*X \setminus \{0\}}
      \frac{\max\{-\II_{\bd X,x}(\xi,\xi),0\}}
      {\abssqr[g_x]{\xi}}.
    \end{equation}
  \end{enumerate}
\end{subequations}
\end{proposition}
\begin{proof}
  \itemref{gaffney.a}~is stated in~\cite[Thm.~3.1.1.1]{grisvard:85}
  (with opposite sign convention for the second fundamental form);
  here we adapted the equality to the case when $\ExtDer[1]\Ww=0$
  (this gives the term $\normsqr[\Sob{\Tensor 2 X}] {\nabla \Ww}$),
  and the boundary term simplifies to the integral over the second
  fundamental form as $\Ww(\normvec{\bd X})=0$ on $\bd X$.

  \itemref{gaffney.b}~now follows from a trace estimate.  Let $Z$ be a
  submanifold of $\bd X$ of dimension $m-1$ such that
  $\II_{\bd Z} \restr {\bd X \setminus Z} \le 0$, and let $X'$ be a
  collar neighbourhood of $Z$ in $X$ of width $a_0>0$; such a
  neighbourhood exists as long as $a_0$ is smaller than the inverse of
  the maximal curvature $\kappa^-_{\bd X}$ on $Z$.  In this case we
  have
  \begin{align*}
    \normsqr[\Lsqr Z] u
    \le a \normsqr[\Lsqr {T^*X'}] {d u}
    + \frac2a\normsqr[\Lsqr{X'}] u.
  \end{align*}
  for all $a \in \interval[open left]0 {a_0}$ and $u \in \Sob{X'}$.
  In particular,
  \begin{align*}
    \normsqr[\Lsqr{\Tensor 2 X}] {\nabla \Ww}
    &=\normsqr[\Lsqr X]{\GradAdj \Ww}
    -  S_{\bd X} (\Ww,\Ww)\\
    &\le \normsqr[\Lsqr X]{\GradAdj \Ww}
     + \kappa^-_{\bd X} \normsqr[\Lsqr{T^*Z}] \Ww\\
    &\le \normsqr[\Lsqr X]{\GradAdj \Ww}
      + \kappa^-_{\bd X} a \normsqr[\Lsqr{\Tensor[!] 2 {X'}}] {\nabla \Ww}
      + \frac{2 \kappa^-_{\bd X}} a
      \normsqr[\Lsqr{T^*X'}] vw\\
    &\le \normsqr[\Lsqr X]{\GradAdj \Ww}
      + \frac12 \normsqr[\Lsqr{\Tensor[!] 2 {X'}}] {\nabla \Ww}
      + 4 (\kappa^-_{\bd X})^2 \normsqr[\Lsqr{T^*X'}] \Ww
  \end{align*}
  using the previous scalar Sobolev trace estimate together with
  \Cor{kato.ineq} and choosing $a=\min\{a_0,1/(2\kappa^-_{\bd X})\}$
  for the last estimate.  Now, we can bring the $1/2$-term on the left
  hand side, and the result follows from multiplying the inequality by
  $2$.
\end{proof}

\subsubsection*{Homothetically scaling spaces}
We now provide some scaling arguments: For a manifold $X$ with
Riemannian metric $g$, we write
\begin{equation}
  \label{eq:eps.homoth}
  \eps X
  \quadtext{for the Riemannian manifold}
  (X, \eps^2g),
\end{equation}
i.e., we change the length scale by a global factor $\eps>0$.  We call
$\eps X$ the \emph{$\eps$-scaled} space.  Here, $\eps$ can be thought
of as a \emph{length scale parameter}, so the integration with respect
to the volume form scale with $\eps^m$, and derivatives scale with
$\eps^{-1}$.  In particular, one can check that the following scaling
properties hold:
\begin{subequations}
  \label{eq:norm.scale}
  \begin{align}
    \label{eq:norm.scale1}
    \normsqr[\Lsqr {\eps X}] \uu
    &=\eps^m \normsqr[\Lsqr X] \uu,
    &\normsqr[\Lsqr {T^*(\eps X)}] \Ww
    &=\eps^{m-2} \normsqr[\Lsqr {T^*X}] \Ww\\
    \label{eq:norm.scale2}
    \normsqr[\Lsqr {T^*(\eps X)}] {\Grad \uu}
    &=\eps^{m-2} \normsqr[\Lsqr {T^*X}] {\Grad \uu},
    &\normsqr[\Lsqr {\eps X}] {\Div_\eps \Ww}
    &=\eps^{m-2} \normsqr[\Lsqr X] {\Div \Ww}\\
    \label{eq:norm.scale3}
    \normsqr[\Lsqr {\Tensor 2 X}] {\nabla \Ww}
    &=\eps^{m-2} \normsqr[\Lsqr {\Tensor 2 X}] {\nabla \Ww},
  \end{align}
\end{subequations}
Here, $\Div_\eps$ is the formal adjoint of $\Grad$ with respect to the
metric $\eps^2 g$.
For the second fundamental form and its integral term we have
\begin{equation}
  \label{eq:2ff.scale}
  \II_{\bd \eps X}= \frac 1\eps \II_{\bd X}
  \quadtext{and}
  S_{\bd \eps X}(\Ww,\Ww)=\eps^{m-2} S_{\bd X}(\Ww,\Ww)
\end{equation}
i.e., the second fundamental form scales as the inverted length, while
the integrated term scales with the factor $\eps^{m-2}$.  Note
that~\eqref{eq:gaffney.a} tells us that all three terms therein for a
homothetic space $\eps X$ scale with the same factor $\eps^{m-2}$, a
key observation for the proof of the next result:
\begin{theorem}[a uniform Gaffney estimate]
  \label{thm:gaffney.scaled}
  Assume that $X_\eps$ is a closed subset of $\R^m$ with
  $\Contspace[2]$-boundary $\bd X_\eps$ such that the second
  fundamental form is non-positive only on a subset
  $Z_\eps:=\bd X_\eps' \cap \bd X_\eps$ of $\bd X_\eps$
  (i.e., $\II_{\bd X_\eps} \restr{Z_\eps} \le 0$) where
  $X_\eps' \subset X_\eps$ is an $\eps$-homothetic closed subset
  ($X_\eps'=\eps X_1'$), then
  \begin{equation}
    \label{eq:gaffney.scaled}
    \normsqr[\Lsqr{\Tensor[!] 2 {X_\eps}}] {\nabla \Ww}
    \le \CGaffney \normsqr[\Hsdom_\eps] \Ww
  \end{equation}
  holds for all $\Ww \in \Hsdom_\eps$, where $\CGaffney$ is as
  in~\eqref{eq:gaffney.c} with $\kappa^-_{\bd X_1'}$ being the
  supremum of the absolute value of the negative principal curvatures
  of the \emph{unscaled} boundary part
  $Z_1 \subset \bd X_1$ of $X_1'$.
\end{theorem}
\begin{proof}
  We have
  \begin{align*}
    \normsqr[\Lsqr{\Tensor[!] 2 {X_\eps}}] {\nabla \Ww}
    &= \normsqr[\Lsqr {X_\eps}]{\GradAdj \Ww}
    - S_{\bd X_\eps}(\Ww,\Ww)\\
    &\le \normsqr[\Lsqr {X_\eps}]{\GradAdj \Ww}
    - S_{Z_\eps}(\Ww,\Ww)
    =\normsqr[\Lsqr {X_\eps}]{\GradAdj \Ww}
    - \eps^{m-2} S_{Z_1} (\Ww,\Ww).
  \end{align*}
  using only the parts of the second fundamental form which are
  positive in the estimate and the scaling
  behaviour~\eqref{eq:2ff.scale}.  Now, we estimate the last unscaled
  boundary integral $S_{Z_1}(\Ww,\Ww)$ as in the proof of
  \Prpenum{gaffney}{gaffney.b} and then use the
  scaling~\eqref{eq:norm.scale2}--\eqref{eq:norm.scale3} in order to
  obtain the desired estimate.
\end{proof}

\subsection{Graph-like spaces, related Hilbert spaces and operators}
\label{ssec:graph-like}

Let us first define graph-like spaces abstractly.  One reason is that
a specific model we have in mind, a smoothened $\eps$-neighbourhood of
a metric graph $\Xzero$ embedded in $\R^m$ will have edge
neighbourhoods $\Xepsed$ \emph{not} of full edge length $\ell_e$ of
the underlying metric edge $\Xzero[e]$, as the vertex neighbourhoods
$\Xepsvx$ need some space, see also \Figs{graph-like-space}{notation}.  We
consider this case as a perturbation of an abstractly defined
graph-like space, where the edge neighbourhoods have full edge length
$\ell_e$, see \Thm{main-b}.

Having the above in mind, we define a family of $\eps$-dependent
spaces $\Xeps$ abstractly as Riemannian manifolds and assume that it
is flat, i.e., the metric as map from a coordinate patch into the
space of positive sesquilinear forms on $TX$ is constant.  In order to
simplify the model, we also assume that the parameter $\eps$ only
enters in the metric, i.e., we think of $\Xeps$ as $(\X,g_\eps)$.  We
have given a similar metric description of a graph-like space
in~\cite[Def.~4.1]{post-simmer:21}:
\begin{definition}[graph-like space]
  \label{def:gs}
  Let $\Xzero$ be a connected metric graph as in \Def{mg}.  We say
  that an $m$-dimensional Riemannian manifold $\Xone$ ($m \ge 2$) is
  an \emph{(abstract) graph-like space} associated with $\Xzero$ if
  \begin{enumerate}
  \item
    \label{gs.a}
    there is a decomposition
    $\Xone = \bigcup_{e \in E} \Xoneed \cup \bigcup_{v \in V}
    \Xonevx$, and
    \begin{equation*}
      \bd_e \Xonevx
      = \bd_v \Xoneed
      := \Xoneed \cap \Xonevx\ne \emptyset
    \end{equation*}
    if and only if $e \in E_v$, and all other sets $\Xoneed$ and
    $\Xonevx$ are disjoint;
  \item
    \label{gs.b}
    each $\Xoneed$ is isometric with $I_e \times \Yoneed$ where
    $I_e=\interval0{\ell_e}$ as in the metric graph and where
    $\Yoneed$ is a compact and connected Riemannian manifold of
    dimension $m-1$, called \emph{(edge) cross sectional} or
    \emph{(edge) transversal} space;
  \item
    \label{gs.c}
    the boundary $\bd \Xone$ is of class $\Contspace[2]$;
  \item
    \label{gs.d}
    as measure on a graph-like space $\Xeps$ we choose the natural
    volume measure of the Riemannian manifold.
  \end{enumerate}
  We denote by $\intr \Xone$ the \emph{interior} of $\Xone$ and by
  $\bd \Xone$ its boundary.
\end{definition}

\begin{definition}[shrinking family of graph-like spaces]
  \label{def:gs.shrinking}
  Let $\Xone$ be a graph-like space associated with a metric graph
  $\Xzero$.  We say that $(\Xeps)_\eps$
  ($\eps \in \interval[open left]01$) is a \emph{shrinking family of
    graph-like spaces} associated with $\Xzero$ if
  \begin{enumerate}
  \item
    \label{gs.shrinking.a}
    each $\Xeps$ is a graph-like space associated with $\Xzero$ in the
    sense of \Def{gs}; we call $\Xone$ the \emph{unscaled} version of
    the family;
  \item
    \label{gs.shrinking.b}
    each vertex neighbourhood $\Xepsvx$ and each cross section
    $\Yepsed$ is $\eps$-homothetic, i.e., $\Xepsvx=\eps\Xonevx$ and
    $\Yepsed=\eps\Yoneed$ (see~\eqref{eq:eps.homoth} for the notation).
  \end{enumerate}
\end{definition}

From the space decomposition (disjoint up to sets of measure zero) we
conclude the following decomposition
\begin{equation}
  \label{eq:gs.decomp}
  \Lsqr {\Xeps}
  = \bigoplus_{e \in E} \Lsqr{\Xepsed} \oplus
  \bigoplus_{v \in V} \Lsqr{\Xepsvx}
  \quadtext{and}
  \Lsqr{\Xepsed}
  \cong \Lsqr{I_e,\Lsqr {\Yepsed}}
  \cong \Lsqr{I_e} \otimes \Lsqr{\Yepsed}
\end{equation}
of the space of functions $\xHS=\Lsqr \Xeps$, so the squared norm
of the latter space can be written as
\begin{equation}
  \label{eq:norm.ed}
  \normsqr[\Lsqr{\Xepsed}] {\uu}
  = \int_{I_e}\normsqr[\Lsqr{\Yepsed}] {\uu_e(s,\cdot)} \dd s.
\end{equation}
For $\uu \in \Lsqr{\Xeps}$, we write $\uu_e$ and $\uu_v$ for the
component on (restriction onto) $\Xepsed$ and $\Xepsvx$.

Let $\Sob {\Xeps}$ be the space of (classes of) functions which are
square-integrable with square-integrable weak derivatives.  As
exterior derivative, we choose
\begin{equation}
  \label{eq:ext.der.gs}
  \map \xGrad {\xhsdom=\Sob {\Xeps}} {\xHs=\Sob[{\ExtDer[1]=0}] {T^*\Xeps}}
  \qquad
  \uu \mapsto \xGrad \uu,
\end{equation}
where $\xGrad \uu$ is the usual exterior derivative with values in the
cotangent bundle $T^*\Xeps$.  Here, the exterior derivative maps into
the closed subspace $\Sob[{\ExtDer[1]=0}] {T^*\Xeps}$ of
$\Lsqr{T^*\Xeps}$, consisting of closed forms (irrotational vector
fields, see~\eqref{eq:1-form.mfd}).

We have
\begin{equation}
  \label{eq:ext.der.ed}
  \xGrad \uu_e = \uu_e' \dd s_e + \xtransvGrad \uu_e,
\end{equation}
where, $\uu_e'$ denotes the derivative with respect to the first
(longitudinal) variable.  Moreover, denote by $\efct[e]$ the
normalised (constant) eigenfunction of the Neumann Laplacian
$(\transv\Grad)^* \transv\Grad$ on $\Yepsed$.

For the $1$-forms, we need a subspace $\xHs$ of
$\Lsqr{T^*\Xeps}$: We also have
\begin{equation}
  \label{eq:gs.ed.dec}
  \Uu_e= \longit \Uu_e \dd s_e + \transv {\Uu}_e
  \qquadtext{where}
  \longit {\Uu}_e := \Uu_e \cdot \dd s_e
\end{equation}
is the longitudinal part of $\Uu_e$.  Moreover, we have
\begin{equation}
  \label{eq:gs.div}
  \xGradAdj \Uu_e= -(\longit \Uu_e)' + \xtransvGradAdj \transv{\Uu}_e.
\end{equation}
On the boundary of $\Xeps$, the normal component of $\Uu$ vanishes; in
particular, on $\Xepsed$, the normal component is $\transv \Uu_e$ and
hence it vanishes.

\begin{lemma}
  \label{lem:ext.der.gs}
  The operator $\xGrad$ defined in~\eqref{eq:ext.der.gs} is an
  abstract exterior derivative in the sense of \Def{ext.der}.  We call
  the associated Dirac-type operator $\xDirac$ with domain
  \begin{align*}
    \xHSdom=\xhsdom \oplus \xHsdom
    =\Sob{\Xeps} \oplus \dom \xGradAdj
    \quadtext{in}
    \xHS=\xhs \oplus \xHs
    = \Lsqr {\Xeps} \oplus \Sob[{\ExtDer[1]=0}] {T^*\Xeps}.
  \end{align*}
  The index of the associated Dirac operator is the Euler
  characteristic of the underlying graph, i.e.,
  $\ind \xDirac=\card V - \card E$.
\end{lemma}
\begin{proof}
  For the index, note that $\Xeps$ is homotopy equivalent with the
  underlying metric graph $\Xzero$, hence their indices
  agree.
\end{proof}


We later need some uniformity assumptions in our analysis; we state it
for the unscaled graph-like space $\Xone$:
\begin{definition}[uniform (family of) graph-like space(s)]
  \label{def:gs.unif}
  We say that a graph-like space $\Xone$ associated with a metric
  graph $\Xzero$ is \emph{uniform} if the underlying metric graph is
  uniform (i.e.~\eqref{eq:ell0} holds) and if
  \begin{subequations}
    \begin{enumerate}
    \item
      \label{gs.unif.a}
      the $(m-1)$-dimensional volume
      \begin{equation}
        \label{eq:vol.y}
        \vol\Yoneed=\vol{\Yone}
      \end{equation}
      is independent of $e \in E$, where $\Yone$ is some
      prototypic\footnote{We make this assumption in order to avoid
        technical complications with \emph{weighted} Kirchhoff
        conditions used e.g.\ in~\cite{exner-post:09}.}  cross
      sectional space;
    \item
      \label{gs.unif.b}
      there is $\tau \in \interval[open left]01$ such that
      $\bd_e \Xonevx$ (isometric with $\Yoneed$ by~\Defenum{gs}{gs.b})
      has an $\tau \ell_e$-collar neighbourhood $\Xonevx[e]$ inside
      $\Xonevx$, i.e., $\Xonevx[e]$ is isometric with
      $\interval0{\tau\ell_e}\times \Yoneed$ (see
      \Fig{notation} for the $\eps$-scaled version);
    \item
      \label{gs.unif.c}
      the inverse isoperimetric constant\footnote{Note that
        $\intbd \Xonevx=\bd \Xonevx \cap \intgS X$ consists of $\deg v$
        many boundary parts of $\bd \Xonevx$ \emph{not} belonging to
        $\bd \Xone$, i.e., we only consider the boundary of
        $\bd \Xone$ in $\intgS X$.  In particular, we have
        $\vol[m-1]{\intbd \Xonevx}=(\deg v)\vol[m-1]{\Yone}$ due
        to~\eqref{eq:vol.y}.}
      \begin{equation}
        \label{eq:rel.vol}
        \Cisoper
        := \sup_{v \in V} \frac{\vol[m]{\Xonevx}}
        {\vol[m-1]{\intbd \Xonevx}} < \infty
      \end{equation}
      of the vertex neighbourhoods is uniformly bounded from above
      (where $\vol[m]\cdot$ denotes the $m$-dimensional Hausdorff
      measure);
    \item
      \label{gs.unif.d}
      we have
      \begin{equation}
        \label{eq:lambda.vx}
        \lambdaVx
        := \inf_{v \in V} \lambda_2(\Delta_{\Xonevx})>0,
      \end{equation}
      where $\lambda_2(\Delta_{\Xonevx})$ is the second (first
      non-zero) eigenvalue of the Neumann Laplacian on $\Xonevx$;
    \item
      \label{gs.unif.e}
      the negative curvature of $\bd \Xonevx$ is uniformly bounded,
      i.e.,
      \begin{equation}
        \label{eq:curv.max}
        \curvMax
        := \sup_{v \in V} \max\{-\II_{\bd \Xonevx},0\}<\infty;
      \end{equation}
    \item
      \label{gs.unif.f}
      we have
      \begin{equation}
        \label{eq:unif.lambda.ed}
        \lambdaEd
        := \inf_{e \in E} \lambda_2(\Delta_{\Yoneed})>0,
      \end{equation}
      where $\lambda_2(\Delta_{\Yoneed})$ is the second (first
      non-zero) eigenvalue of the Neumann Laplacian on $\Yoneed$.
    \end{enumerate}
  \end{subequations}
  We say that a shrinking family of graph-like spaces $(\Xeps)_{\eps \in \interval[open left]01}$
  associated with $\Xzero$ is \emph{uniform} if the unscaled version
  $\Xone$ of the graph-like family is uniform in the above sense.
\end{definition}

\subsubsection*{Some estimates on uniform families of shrinking
  graph-like spaces}
We now state some results from~\cite[Ch.~6]{post:12}.  We first use
the following decomposition for finite length edges: Denote by
$\Xepsed[,v]$ the subset of $\Xepsed$ isometric with
$\interval 0 {\ell_e/2} \times \Yepsed$ touching $\Xepsvx$ (we cut the
edge neighbourhood $\Xepsed$ into two halves).  For semi-infinite
edges we just set $\Xepsed[,v]=\Xepsed$. Then we have the
decomposition
\begin{equation}
  \label{eq:enlarged.vx.nbhd}
  \Xeps = \bigcup_{v \in V} \starXepsvx,
  \quadtext{where}
  \starXepsvx = \Xepsvx \cup \bigcup_{e \in E_v} \Xepsed[,v]
\end{equation}
into star-shaped \emph{enlarged} vertex neighbourhoods $\starXepsvx$;
we also set $u_{\stSh v}:= u \restr \starXepsvx$.

For a Riemannian manifold $M$, we denote by
\begin{equation}
  \label{eq:dashint}
  \dashint_M \uu=\frac1{\vol M} \int_M \uu \dd M
\end{equation}
the \emph{average value} of $u$ on $M$.  Note that for
$\eps$-homothetic manifolds we have
$\dashint_{\eps M} \uu = \dashint_M \uu$ (used for the two average
values in the following lemma):

The next lemma is~\cite[Prop.~5.1.3]{post:12}:
\begin{lemma}
  \label{lem:est.vx.av}
  We have
  \begin{align*}
    \sum_{v \in V} \sum_{e \in E_v}
      \vol{\bd_e \Xepsvx}
      \Bigabssqr{\dashint_{\bd_e \Xonevx} \uu_v - \dashint_{\Xonevx} \uu_v}
    &\le \eps \sum_{v \in V}
      \Bigl(\tau \min\{\ell_0,1\} + \frac 2{\tau \ell_0 \lambda_2(\Xonevx)}
      \Bigr) \normsqr[\Lsqr{T^* \Xepsvx}] {\xGrad\uu_v}\\
    &\le \eps \Cvxcol \sum_{v \in V} \normsqr[\Lsqr{T^* \Xepsvx}] {\xGrad\uu_v},
  \end{align*}
  where
  \begin{equation}
    \label{eq:def.cvxcol}
    \Cvxcol := \tau + \frac 2{\tau \ell_0 \lambdaVx}.
  \end{equation}
\end{lemma}

A proof of the next lemma can be found in~\cite[Lem.~6.3.6]{post:12}
(using here alternatively the optimal estimate for the Sobolev trace
estimate $\Sob{\Xepsed[,v]} \to \Lsqr{\bd_v \Xepsed}$; the squared
norm of the corresponding operator is $\coth(\ell_e/2)$).
\begin{lemma}
  \label{lem:vx.est}
  For a uniformly shrinking family of graph-like spaces we have
  \begin{subequations}
    \begin{align*}
      \normsqr[\Lsqr{\Xepsvx}]{u_v}
      \le \Cvx(v,\eps) \normsqr[\Sob \starXepsvx] {u_{\stSh v}},
    \end{align*}
    for all $u \in \Sob \starXepsvx$, where
    \begin{multline}
      \label{eq:vx.est}
      \Cvx(v,\eps)
      := 4\max_{e\in E_v}
      \Bigl(\frac{\eps^2}{\lambda_2(\Xonevx)}
      + \frac{\eps^2\vol[m]{\Xonevx}}{\vol[m-1]{\intbd \Xonevx}}
      \Bigl(\tau\min\{\ell_e,1\} + \frac 2{\tau \ell_e \lambda_2(\Xonevx)}
      \Bigr)\\
      +\frac{\eps \vol[m]{\Xonevx}}{\vol[m-1]{\intbd \Xonevx}}\coth(\ell_e/2)
      \Bigr)
    \end{multline}
    Moreover, $\Cvx(v,\eps)\le \eps \Cvx$ for $\eps \in \interval[open left]01$ where
    \begin{equation}
      \label{eq:def.cvx}
      \Cvx :=  4\Bigl(\frac1{\lambdaVx}
      + \Cisoper \bigl(\Cvxcol + \coth(\ell_0/2)\bigr) \Bigr)
    \end{equation}
  \end{subequations}
\end{lemma}
Note that the convergence rate of $\Cvx(\eps,v)$ in \Lem{vx.est}
cannot be better than $\eps$ as $u_{\stSh v}=\1_{\starXepsvx}$ shows:
\begin{align*}
  \frac {\normsqr[\Lsqr{\Xepsvx}]{u_v}}
  {\normsqr[\Sob{\starXepsvx}] u}
  = \frac{\vol[m]{\Xepsvx}}
    {\vol[m]{\Xepsvx}+\sum_{e \in E_v} \vol[m]{\Xepsed[,v]}}
  = \frac\eps{\eps + \sum_{e \in E_v} \ell_e/(2\vol[m]{\Xonevx})}.
\end{align*}


The next lemma follows easily by noting that
$\tuu_e-\iprod[\Lsqr{\Yepsed}]{\tuu_e}{\efct[e]} \efct[e]$ is the
projection onto the orthogonal complement of $\efct[e]$ on the
cross sectional space:
\begin{lemma}
  \label{lem:ed.est}
  For a uniformly shrinking family of graph-like spaces we have
  \begin{align*}
    \int_{I_e} \bignormsqr[\Lsqr{\Yepsed}]
    {\uu_e(s,\cdot)-\iprod[\Lsqr{\Yepsed}]{\uu_e(s,\cdot)}{\efct[e]} \efct[e]}
    \dd s
    \le \frac {\eps^2}{\lambdaEd}
    \int_{I_e} \bignormsqr[\Lsqr{T^*\Yepsed}]
    {\xtransvGrad \uu_e(s,\cdot)}
    \dd s
  \end{align*}
  for all $u \in \Sob{\Xepsed}$.
\end{lemma}

\begin{definition}[convex graph-like space]
  \label{def:gs.convex}
  We say that a graph-like space $\Xone$ is \emph{convex} if
  \begin{enumerate}
  \item
    \label{gs.convex.a}
    each cross-sectional boundary $\bd \Yoneed$ is convex in
    $\Yoneed$, i.e., the second fundamental form fulfils
    $\II_{\bd \Yoneed} \ge 0$ for all $e \in E$;
  \item
    \label{gs.convex.b}
    and if each vertex neighbourhood $\Xonevx$ is simply connected
    for all $v \in V$.
  \end{enumerate}
\end{definition}

\begin{lemma}
  \label{lem:trans.ed}
  For a convex uniformly shrinking family of graph-like spaces we have
  \begin{equation*}
    \normsqr[\Lsqr{T^*\Yepsed}] {\tUu}
    \le \frac {\eps^2(m-1)}\lambdaEd
    \normsqr[\Lsqr{\Tensor[!] 2 \Yepsed}] {\transv \nabla \tUu}
  \end{equation*}
  where $\transv \nabla$ is the covariant derivative on the bundle
  $T^*\Yepsed$.
\end{lemma}
\begin{proof}
  Note that if $\bd \Yoneed$ is convex then $\Yoneed$ is convex and
  hence simply connected.  In particular,
  $\dim \ker \xtransvGradAdj=\{0\}$ and therefore also
  $\dim \ker (\xtransvGrad\xtransvGradAdj)=\{0\}$.  Moreover,
  \begin{align*}
    0 \notin \spec{\xtransvGrad\xtransvGradAdj}
    \qquadtext{and}
    \inf \spec{\xtransvGrad\xtransvGradAdj}
    =\lambda_2(\Yepsed)
    =\frac{\lambda_2(\Yoneed)}{\eps^2}>0
  \end{align*}
  as $\Yepsed=\eps\Yoneed$ is compact and connected.  From the min-max
  characterisation of eigenvalues we conclude
  \begin{equation}
    \label{eq:lambda.ed}
    \normsqr[\Lsqr{T^*\Yepsed}] {\tUu}
    \le \frac 1{\inf \spec{\xtransvGrad\xtransvGradAdj}}
    \normsqr[\Lsqr\Yepsed] {\xtransvGradAdj \tUu}
    \le \frac {\eps^2(m-1)}
    {\lambdaEd} \normsqr[\Lsqr{\Tensor[!] 2 \Yepsed}] {\transv \nabla \tUu}
  \end{equation}
  together with \Lem{triv.emb}.
\end{proof}
For the estimate in \Lem{trans.ed} it suffices that $\Yepsed$ is
simply connected, so that $\dim \ker \xtransvGradAdj=\{0\}$.

\subsubsection*{Graph-like spaces arising as neighbourhoods of embedded
  metric graphs}

We now consider a ``real world'' example as a perturbation of our
abstract graph-like space, see also \Figs{graph-like-space}{notation}:
\begin{definition}[embedded graph-like space]
  \label{def:gs.emb}
  We say that $\wtXone$ is an \emph{embedded (uniform) graph-like
    space} if the following holds:
  \begin{enumerate}
  \item
    \label{gs.emb.a}
    the underlying metric graph $\Xzero$ is embedded in $\R^m$ such
    that its edges correspond to straight line segments in $\R^m$;

  \item
    \label{gs.emb.b}
    we require \Defs{gs}{gs.unif} to hold (with the obvious notation
    for the vertex and edge neighbourhoods) except for
    \Defenum{gs}{gs.b}: each edge neighbourhood $\wtXoneed$ is
    isometric with $\interval 0 {\ell_e(1-\tau)}$ for some
    $\tau \in \interval[open left]01$;

  \item
    \label{gs.emb.c}
    we assume that the boundary $\bd \wt X$ on the vertex
    neighbourhood $\wtXonevx$ is of class $\Contspace[2]$.%
  \end{enumerate}
  We say that a uniform family of graph-like spaces
  $(\wtXeps)_{\eps \in \interval[open left]01}$ is \emph{embedded} if
  $\wtXone$ is embedded in the sense above and if additionally, the
  conditions of \Def{gs.shrinking} hold.  In this case $\tau$
  becomes $\eps\tau$.
\end{definition}
The condition in \Defenum{gs.emb}{gs.emb.c} is already included in
\Defenum{gs}{gs.c}, but we repeat it here as if one starts with an
$\eps/2$-neighbourhood of an embedded metric graph $\Xzero$, then a
vertex neighbourhood of a vertex of degree larger than $2$ will have
non-convex non-smooth parts.


%
\section{Proof of the main results}
\label{sec:proofs}
%

We now prepare the proof of \Thm{main-a}, i.e., we show that $\mDirac$
and $\xDirac$ are $\delta$-quasi unitarily equivalent~(see \Def{que}).
\subsection{Identification operators for \Thm{main-a}}
\label{ssec:id-ops}

We now define
\begin{equation}
  \label{eq:id-op}
  \map{\IdOp}{\mHS=\mhs \oplus \mHs}{\xHS=\xhs \oplus \xHs} \qquad
  \FF = (\ff,\Ff) \mapsto
  \IdOp \FF = (\idop \ff, \Idop \Ff).
\end{equation}
For brevity, we write $\IdOp=\idop \oplus \Idop$.  We define
\begin{subequations}
  \begin{align}
    \label{eq:id.op.0}
    (\idop \ff)_e:= \ff_e \otimes \efct[e],
    \qquad
    (\idop \ff)_v:= 0,\\
    \label{eq:id.op.1}
    (\Idop \Ff)_e:= \Ff_e \otimes \efct[e],
    \qquad
    (\Idop \Ff)_v:= 0.
  \end{align}
\end{subequations}
Here, we think of $\Ff_e$ as having $\dd s_e$ already inside, and we
interpret $\dd s_e$ also as an element of $T^*\Xepsed$.  Then we have
\begin{subequations}
  \begin{align}
    \nonumber
    \iprod[\xhs]{\idop \ff}{\uu}
    &= \sum_{e \in E}
      \int_{I_e} \ff_e(s) \iprod[\Lsqr {\Yepsed}]{\efct[e]}{\uu_e(s,\cdot)} \dd s
      =\iprod[\mhs]{\ff}{\idopAdj\uu},
    \\
    \label{eq:fct.j.adj}
    (\idopAdj \uu)_e(s)
    &=\iprod[\Lsqr{\Yepsed}]{\uu_e(s,\cdot)}{\efct[e]},\\ 
    \nonumber
    \iprod[\mHs]{\Idop \Ff}{\Uu}
    &= \sum_{e \in E}
      \int_{I_e} \Ff_e(s)\cdot \dd s_e
      \iprod[\Lsqr {\Yepsed}]{\efct[e]}{\Uu_e(s,\cdot) \cdot \dd s_e} \dd s\\
    \nonumber
    &= \sum_{e \in E}
      \int_{I_e} \Ff_e(s) \cdot
      \iprod[\Lsqr {\Yepsed}]{\longit{\Uu}_e(s,\cdot)}{\efct[e]}\dd s_e \dd s
      =\iprod[\mHs]{\Ff}{\IdopAdj\Uu},
    \\
    \label{eq:vct.j.adj}
    (\IdopAdj \Uu)_e(s)
    &=\iprod[\Lsqr{\Yepsed}]{\longit{\Uu}_e(s,\cdot)}{\efct[e]}
  \end{align}

\end{subequations}
for any $\UU=(\uu,\Uu) \in \xHS$ using~\eqref{eq:gs.ed.dec}.  It is easily seen
that
\begin{equation}
  \label{eq:j-star.j.id}
  \idopAdj \idop = \id \mhs, \qquad
  \IdopAdj \Idop = \id \mHs, \quadtext{hence}
  \IdOpAdj \IdOp = \id \mHS.
\end{equation}

\begin{lemma}
  \label{lem:fct.que.j}
  For the function identification operator, we have
  \begin{align*}
    \normsqr[\Lsqr \Xeps]{\uu - \idop \idopAdj\uu}
    &\le  \Bigl(\eps \Cvx + \frac{\eps^2}{\lambdaEd}\Bigr)
      \normsqr[\Sob \Xeps] \uu
  \end{align*}
  for all $\uu \in \Sob \Xeps$.
\end{lemma}
\begin{proof}
  We have
  \begin{align*}
    \normsqr[\Lsqr \Xeps]{\uu - \idop \idopAdj\uu}
    &= \sum_{v \in V} \normsqr[\Lsqr {\Xepsvx}] {\uu_v}
      + \sum_{e \in E} \int_{I_e} \bignormsqr[\Lsqr{\Yepsed}]
      {\uu_e(s,\cdot)-\iprod[\Lsqr{\Yepsed}]{\uu_e(s,\cdot)}{\efct[e]} \efct[e]}
      \dd s\\
    &\le \eps \Cvx \sum_{v \in V} \normsqr[\Sob \starXepsvx] {u_{\stSh v}}
      +\frac {\eps^2}{\lambdaEd} \sum_{e \in E} \normsqr[\Lsqr{T^*\Xepsed}] {\xGrad u_e}\\
    &\le \Bigl(\eps \Cvx+\frac {\eps^2}{\lambdaEd}\Bigr)
      \normsqr[\Sob \Xeps]  u
  \end{align*}
  by \Lems{vx.est}{ed.est}.  Note that
  $\int_{I_e} \bignormsqr[\Lsqr{T^*\Yepsed}] {\xtransvGrad
    \uu_e(s,\cdot)} \dd s \le \normsqr[\Lsqr{T^*\Xepsed}] {\xGrad
    \uu_e}$.
\end{proof}

\begin{lemma}
  \label{lem:vct.que.j}
  For the $1$-form identification operator, we have
  \begin{align*}
    \normsqr[\Lsqr {T^*\Xeps}]{\Uu - \Idop \IdopAdj\Uu}
    &\le  \Bigl(\eps \Cvx + \frac{m\eps^2}{\lambdaEd}\Bigr)
      \CGaffney
      \bigl(\normsqr[\Lsqr {T^*\Xeps}] \Uu
      + \normsqr[\Lsqr \Xeps] {\xGradAdj \Uu}\bigr)
  \end{align*}
  for all $\Uu \in \Sob{T^*\Xeps}=\dom \xGradAdj=\xHsdom$.
\end{lemma}
\begin{proof}
  \begin{align}
    \nonumber
    \normsqr[\Lsqr {T^*\Xeps}]{\Uu - \Idop \IdopAdj\Uu}
    &= \sum_{v \in V} \normsqr[\Lsqr {\Xepsvx}] {\Uu_v}
      + \sum_{e \in E} \int_{I_e} \bignormsqr[\Lsqr{\Yepsed}]
      {\longit{\Uu}_e(s,\cdot)
      -\iprod[\Lsqr{\Yepsed}]{\longit{\Uu}_e(s,\cdot)}{\efct[e]} \efct[e]}
      \dd s\\
    \label{eq:vct.que.j}
    & \hspace*{0.2\textwidth}
      + \sum_{e \in E} \int_{I_e} \bignormsqr[\Lsqr{\Yepsed}]
      {\transv{\Uu}_e(s,\cdot)} \dd s.
  \end{align}
  For the first term, we use the $0$-form estimate, \Lem{vx.est},
  together with Kato's inequality \Cor{kato.ineq} and obtain
  \begin{equation*}
    \sum_{v \in V} \normsqr[\Lsqr {\Xepsvx}] {\Uu_v}
    \le \eps\Cvx\sum_{v \in V} \normsqr[\Sob{\Tensor[!] 2 \starXepsvx}]
        {\Uu_{\stSh v}}
    = \eps\Cvx \normsqr[\Sob{\Tensor[!] 2 \Xeps}] \Uu
  \end{equation*}
  (recall the definition of the enlarged vertex
  neighbourhood~\eqref{eq:enlarged.vx.nbhd}).

  For the second term, we use the $0$-form estimate \Lem{ed.est},
  together with \Lem{kato.ineq} and obtain
  \begin{align*}
    \sum_{e \in E} \int_{I_e} \bignormsqr[\Lsqr{\Yepsed}]
      {\longit{\Uu}_e(s,\cdot)
      -\iprod[\Lsqr{\Yepsed}]{\longit{\Uu}_e(s,\cdot)}{\efct[e]} \efct[e]}
      \dd s
    &\le \frac{\eps^2} \lambdaEd
      \sum_{e \in E} \int_{I_e} \normsqr[\Lsqr{\Tensor[!] 2 \Yepsed}]
        {\transv \nabla \Uu_e(s,\cdot)} \dd s\\
    &\le \frac{\eps^2} \lambdaEd
      \normsqr[\Lsqr{\Tensor[!] 2 \Yepsed}] {\nabla \Uu},
  \end{align*}
  For the third term we use \Lem{trans.ed} and obtain
  \begin{align*}
    \sum_{e \in E} \int_{I_e} \bignormsqr[\Lsqr{\Yepsed}]
    {\transv \Uu_e(s,\cdot)} \dd s
    & \le \frac {\eps^2(m-1)}\lambdaEd
      \sum_{e \in E} \int_{I_e}
      \normsqr[\Lsqr{\Tensor[!] 2 \Yepsed}] {\transv \nabla \Uu(s,\cdot)} \dd s\\
    & \le \frac {\eps^2(m-1)}\lambdaEd
      \normsqr[\Lsqr{\Tensor[!] 2 \Xeps}] {\nabla \Uu}.
  \end{align*}
  Summing the three contributions we get the first factor of the right
  hand side of~\eqref{eq:vct.que.j}.  Finally, note that for a
  uniformly shrinking family of graph-like space the uniform Gaffney
  estimate can be used: only the vertex neighbourhoods
  \begin{align*}
    \Xeps'
    = \bigcup_{v \in V} \Xepsvx
  \end{align*}
  have negative principal curvatures on $\bd \Xeps' \cap \bd \Xeps$,
  the principal curvature on $\Xeps \setminus \Xeps'$ is given by the
  curvature of $\Yepsed$ and hence non-negative (cf.\ \Lem{2ff.prod}).
  In particular, we can apply \Thm{gaffney.scaled}.
\end{proof}

\begin{remark}
  Note that $\Idop$ maps into the correct space.  If there were any
  harmonic form $\Uu$ arising from a non-simply-connected vertex
  neighbourhood $\Xepsvx$ (note that we have excluded this case in
  \Def{gs.convex}), then any cross-sectional integral over
  $(\{s\} \times \Yepsed \subset \Xepsed)$ on an edge neighbourhood
  vanishes: Note first that such an integral depends only on the
  homology class of $(\{s\} \times \Yepsed, \{s\} \times \bd \Yepsed)$
  in $(\Xeps,\bd \Xeps)$ (see e.g.~\cite{cdtg:02} for the
  three-dimensional case).  Second, for a harmonic form $\Uu$
  enclosing a hole in $\Xepsvx$, the flux through the internal
  boundary $\intbd \Xepsvx$ vanishes, hence all cross-sectional
  integrals as above vanish also.  In particular, $\Idop \Ff$ is
  orthogonal to $\Uu$, as
  \begin{align*}
    \iprod[\Lsqr{T^*\Xeps}] {\IdOp \Ff}\Uu
    &=\sum \int_{I_\eps}
    \Ff_e(s) \cdot \dd s_e
    \iprod[\Lsqr \Yepsed]{\efct[e]}{\Uu_e(s,\cdot)\ddot \dd s_e}
    \dd s_e\\
    &=\sum \int_{I_\eps}
    \Ff_e(s) \efct \int_{\Yepsed} \longit \Uu_e(s,\cdot) \dd s_e
    =0,
  \end{align*}
  as $\int_{\Yepsed} \longit \Uu_e(s,\cdot) \dd s_e$ is the flux
  through $\{s\}\times \Yepsed$ of $\Uu$.
\end{remark}

The operators of interest are the resolvents
\begin{equation}
  \label{eq:resovents}
  \mRes^\pm:=(\mDirac \mp \im)^{-1} \qquadtext{and}
  \xRes^\pm:=(\xDirac \mp \im)^{-1}.
\end{equation}

Denote by $\lsqr{V,\deg}$ the Hilbert space of functions $\map a V \C$ such that
\begin{equation}
  \label{eq:lsqr.v.deg}
  \normsqr[\lsqr{V,\deg}] a
  :=  \sum_{v \in V} \abssqr{a(v)} \deg v < \infty.
\end{equation}
\begin{lemma}
  \label{lem:res.diff}
  We have
  \begin{align*}
    \xRes^\pm \IdOp -\IdOp \mRes^\pm
    = \xOpA^*\mOpA + \xOpB^*\mOpB,
  \end{align*}
  where
  \begin{align*}
    \map {&\mOpA} {\mHS}{\hsaux:=\lsqr{V,\deg}},
    &&\mOpA \GG := (\ff(v))_{v \in V},\\
    \map {&\mOpB} {\mHS}{\Hsaux:=\bigoplus_{v \in V}\C^{E_v}},
    &&\mOpB \GG := \bigl((\Ff_e(v))_{e \in E_v}\bigr)_{v \in V}
  \end{align*}
  for $(\ff,\Ff)=\FF=\mRes^\pm \GG$ and
  \begin{align*}
    \map {&\xOpA} {\xHS}{\hsaux},
    &&\xOpA \WW := \Bigl(\frac1{\deg v} \iprod[\xhsvx]{\xGradAdj \Uu}{\efct[e]}\Bigr)_{v \in V}\\
    \map {&\xOpB} {\xHS}{\Hsaux},
    &&\xOpB \WW := \bigl((\iprod[\Lsqr{\bd_e \Xepsvx}]{\uu_v}
       {\efct[e]}-C_v \uu)_{e \in E_v}\bigr)_{v \in V}
  \end{align*}
  for $(\uu,\Uu)=\UU=\xRes^\mp \WW$, where
  \begin{align*}
    C_v \uu
    = \Bigl(\frac{\vol{\Yone}}{\eps\vol{\Xonevx}}\Bigr)^{1/2}
     \iprod[\Lsqr{\Xepsvx}] \uu {\efct[v]}
    =  (\eps^{m-1}\vol{\Yone})^{1/2}\dashint_{\Xonevx} \uu.
  \end{align*}
  Here, $\dashint_{\Xepsvx} \uu$ is the average value of $\uu$ on
  $\Xepsvx$ (defined in~\eqref{eq:dashint}) and and $\efct[v]$ is the
  normalised constant eigenfunction on $\Xepsvx$ (with value
  $\vol{\Xepsvx}^{-1/2}$).
\end{lemma}
\begin{proof}
  For $\FF=\mRes^\pm \GG$ and $\UU=\xRes^\mp \WW$, we have
  \begin{align*}
    \iprod[\xHS]{(\IdOp \mRes^\pm-\xRes^\pm \IdOp)\GG}{\WW}
    &=\iprod[\xHS]{\IdOp \FF}{\xDirac \UU}
      -\iprod[\xHS]{\IdOp \mDirac \FF}{\UU}\\
    &=\bigl(\iprod[\xhs]{\idop \ff}{\xGradAdj \Uu}
      - \iprod[\xHs]{\Idop \mGrad \ff}{\Uu}\bigr)
      + \bigl(\iprod[\xHs]{\Idop \Ff}{\xGrad \uu}
      - \iprod[\xhs]{\idop \mGradAdj \Ff}{\uu} \bigr).
  \end{align*}
  For the first difference, we have
  \begin{align*}
    \iprod[\xhs]{\idop \ff}{\xGradAdj \Uu}
    - \iprod[\xHs]{\Idop \mGrad \ff}{\Uu}
    &=
      \sum_{e \in E}\bigl(
      \iprod[\xhsed]{\ff_e \otimes \efct[e]}{\xGradAdj \Uu_e}
      -\iprod[\xHsed]{\ff'_e \otimes \efct[e] \dd s_e}{\Uu_e}
      \bigr)\\
    &=-\sum_{e \in E}
      \iprod[\Lsqr{\bd \Xepsed}]{\ff_e \otimes \efct[e]}
      {\Uu_e \cdot \normvec{\bd \Xepsed}}\\
    &=\sum_{v \in V} \sum_{e \in E_v} \ff_e(v) \iprod[\Lsqr{\bd_e \Xepsvx}] \efct
      {\Uu_v \cdot \normvec{\bd_e \Xepsvx}}\\
    &=\sum_{v \in V} \ff(v) \iprod[\Lsqr{\bd \Xepsvx}] \efct
      {\Uu_v \cdot \normvec{\bd \Xepsvx}}\\
    &=\sum_{v \in V} \ff(v) \iprod[\Lsqr{\Xepsvx}] \efct{\xGradAdj \Uu_v}
      =\iprod[\hsaux]{\mOpA \GG}{\xOpA \WW}
  \end{align*}
  using $(\mGrad \ff)_e=\ff'_e \dd s_e$ for the first equality,
  integration by parts~\eqref{eq:part.int} and
  $\xGrad(\ff_e \otimes \efct[e])=\ff'_e \otimes \efct[e] \dd s_e$ for
  the second, a reordering and the fact that
  $\Uu \cdot \normvec{\bd \Xeps}=0$ for the third, $\ff_e(v)=\ff(v)$
  for the fourth and the divergence theorem for the last equality.  We
  have also employed the assumption that $\vol{\Yoneed}=\vol{\Yone}$
  is independent of $e \in E$, hence $\efct[e]=\efct$ has a common
  value by~\eqref{eq:vol.y}.

  For the second difference, we have
  \begin{align*}
    \iprod[\xHs]{\Idop \Ff}{\xGrad \uu}
    - \iprod[\xhs]{\idop \mGradAdj \Ff}{\uu}
    &=
      \sum_{e \in E}\bigl(
      \iprod[\xHsed]{\Ff_e \otimes \efct[e]}{\xGrad \uu_e}
      -\iprod[\xhsed]{-\Ff'_e \cdot \dd s_e \otimes \efct[e]}{\uu_e}
      \bigr)\\
    &=\sum_{e \in E}
      \iprod[\Lsqr{\bd \Xepsed}]{(\Ff_e \cdot \dd s_e) \otimes \efct[e] \dd s_e
      \cdot \normvec{\bd \Xepsed}} {\uu_e}\\
    &=\sum_{v \in V} \sum_{e \in E_v} \Ff_e(v)
      \iprod[\Lsqr{\bd_e \Xepsvx}] {\efct[e]}
      {\uu_v}\\
    &=\sum_{v \in V} \sum_{e \in E_v} \Ff_e(v)
      \bigl(\iprod[\Lsqr{\bd_e \Xepsvx}] {\efct[e]} {\uu_v}
      -\conj{C_v \uu} \bigr)\\
    &= \iprod[\Hsaux]{\mOpB \GG}{\xOpB \WW}
  \end{align*}
  using $(\mGradAdj \Ff)_e =-\Ff'_e \cdot \dd s_e$ for the first
  equality, integration by parts~\eqref{eq:part.int} again and
  $\xGrad(\ff_e \otimes \efct[e])=\ff'_e \otimes \efct[e] \dd s_e$ for
  the second, a reordering and the sign convention of the oriented
  evaluation~\eqref{eq:or.eval} for the third,
  $\sum_{e \in E_v} \Ff_e(v)=0$ for the fourth.
\end{proof}

\begin{lemma}
  \label{lem:est1}
  The operators $\mOpA$ and $\mOpB$ introduced in \Lem{res.diff}
  fulfil $\normsqr{\mOpA} \le \coth(\ell_0/2)$ and
  $\normsqr{\mOpB} \le \coth(\ell_0/2)$.
\end{lemma}
\begin{proof}
  We estimate
  \begin{align*}
    \normsqr[\lsqr{V,\deg}] {\mOpA \GG}
    = \sum_{v \in V} \abssqr{\ff(v)}\deg v
    &=\sum_{v \in V} \sum_{e\in E_v} \abssqr{\ff_e(v)}
    =\sum_{e \in E}
    \bigl(\abssqr{\ff_e(0)} + \abssqr{\ff_e(\ell_e)} \bigr)\\
    &\le \sum_{v \in V} \coth(\ell_e/2) \normsqr[\Sob{I_e}]{\ff_e}
    \le \coth(\ell_0/2) \normsqr[\Sob{\Xzero}] \ff,
  \end{align*}
  where we used a reordering in the third equality and an optimal
  trace estimate in the first inequality (see
  e.g.~\cite[Sec.~6.1]{post:16}).  Moreover,
  \begin{align*}
    \normsqr[\Sob{\Xzero}] \ff
    = \normsqr[\mhs]{\ff} + \normsqr[\mHs]{\mGrad \ff}
    \le \normsqr[\mHSdom] \ff
    =\normsqr[\mHS] \GG
  \end{align*}
  using~\eqref{eq:norm.dirac.b}
The argument for $\mOpB$ is similar.
\end{proof}

\begin{lemma}
  \label{lem:est2}
  The operator $\xOpA$ defined in \Lem{res.diff} fulfils
  \begin{align*}
    \normsqr{\xOpA} \le \eps\Cisoper,
  \end{align*}
  where the inverse isoperimetric constant is defined
  in~\eqref{eq:rel.vol}.
\end{lemma}
\begin{proof}
  We estimate
  \begin{align*}
    \normsqr[\lsqr{V,\deg}] {\xOpA \WW}
    = \sum_{v \in V} \frac1{\deg v}
    \bigabssqr{\iprod[\xhsvx]{\xGradAdj \Uu}{\efct[e]}}
    \le \sum_{v \in V} \frac{\normsqr[\Lsqr{\Xepsvx}]{\efct[e]}} {\deg v}
    \normsqr[\Lsqr{\Xepsvx}] {\xGradAdj \Uu}
  \end{align*}
  using Cauchy-Schwarz.  The result follows from the fact that
  $\normsqr[\Lsqr{\Xepsvx}]{\efct[e]}=\vol{\Xepsvx}/\vol{\Yepsed}$
  and~\eqref{eq:vol.y} together with
  \begin{align*}
    \normsqr[\Lsqr{\Xeps}]{\xGradAdj \Uu}
    =\normsqr[\xhs]{\xGradAdj \Uu}
    \le \normsqr[\xHSdom] \UU
    =\normsqr[\xHS] \WW
  \end{align*}
  using~\eqref{eq:norm.dirac.b}
\end{proof}

\begin{lemma}
  \label{lem:est3}
  The operator $\xOpB$ defined in \Lem{res.diff} fulfils
  \begin{align*}
    \normsqr{\xOpB}
    \le \eps \Cvxcol,
  \end{align*}
  where $\Cvxcol$ is defined in~\eqref{eq:def.cvxcol}.
\end{lemma}
\begin{proof}
  We estimate
  \begin{align*}
    \normsqr[\Hsaux] {\xOpB \WW}
    = \sum_{v \in V} \sum_{e \in E_v}
      \bigabssqr{\iprod[\Lsqr{\bd_e \Xepsvx}]{\uu_v}{\efct[e]}-C_v \uu}
    &= \sum_{v \in V} \sum_{e \in E_v}
      \vol{\bd_e \Xepsvx}
      \Bigabssqr{\dashint_{\bd_e \Xepsvx} \uu_v - \dashint_{\Xepsvx} \uu_v}\\
    &\le \eps \Cvxcol
      \sum_{v \in V} \normsqr[\Lsqr{T^* \Xepsvx}] {\xGrad\uu_v}
  \end{align*}
  using~\Lem{est.vx.av}.  The result follows from
  \begin{align*}
    \normsqr[\Lsqr{T^* \Xeps}] {\xGrad\uu}
    =\normsqr[\xHs]{\xGrad \uu}
    \le \normsqr[\xHSdom] \UU
    =\normsqr[\xHS] \WW
  \end{align*}
  using again~\eqref{eq:norm.dirac.b}.
\end{proof}

\begin{proof}[Proof of \Thm{main-a}]
  First, the identification operator $\IdOp$ is an isometry, hence the
  first condition of~\eqref{eq:que.1} is fulfilled with $\delta=0$.
  For the second one, note that
  \begin{align*}
    \normsqr[\xHS]{(\id \xHS-\IdOp\IdOpAdj)\xRes^\pm \WW}
    &= \normsqr[\xHS]{(\id \xHS-\IdOp\IdOpAdj)\UU}\\
    &=\normsqr[\xhs]{(\id \xhs-\idop\idopAdj)\uu}
    +\normsqr[\xHs]{(\id \xHs-\Idop\IdopAdj)\Uu}\\
    &\le \Bigl(\eps \Cvx + \frac{\eps^2}{\lambdaEd}\Bigr)
      \normsqr[\xhsdom] \uu
    +\Bigl(\eps \Cvx + \frac{m\eps^2}{\lambdaEd}\Bigr)
      \CGaffney \normsqr[\xHsdom] \Uu\\
    &\le \Bigl(\eps \Cvx + \frac{m\eps^2}{\lambdaEd}\Bigr)
      \CGaffney \normsqr[\xHSdom] \UU
    = \Bigl(\eps \Cvx + \frac{m\eps^2}{\lambdaEd}\Bigr)
      \CGaffney \normsqr[\xHS] \WW
  \end{align*}
  for $(\uu,\Uu)=\UU=\xRes^\pm \WW$ using \Lems{fct.que.j}{vct.que.j}
  together with~\eqref{eq:norm.dirac.a} for the last equality.  Note
  that
  \begin{align*}
    \normsqr[\Sob X] \uu=\normsqr[\xhsdom] \uu
    \qquadtext{and}
    \normsqr[\Lsqr {T^*\Xeps}] \Uu
    + \normsqr[\Lsqr \Xeps] {\xGradAdj \Uu}
    =\normsqr[\xHsdom] \Uu;
  \end{align*}
  moreover $\CGaffney \ge 1$.

  For the resolvent difference~\eqref{eq:que.2} we use
  \LemS{res.diff}{est3} and obtain
  \begin{align*}
    \norm[\Lin{\mHS,\xHS}]{\xRes^\pm \IdOp -\IdOp \mRes^\pm}
    &\le \norm[\Lin{\mHS,\xHS}]{\xOpA^*\mOpA}
    + \norm[\Lin{\mHS,\xHS}]{\xOpB^*\mOpB}\\
    &= \norm[\Lin{\xHS,\hsaux}] \xOpA
      \norm[\Lin{\mHS,\hsaux}] \mOpA
    + \norm[\Lin{\xHS,\Hsaux}] \xOpB
      \norm[\Lin{\mHS,\Hsaux}] \mOpB\\
    &\le \eps\bigl(\Cisoper + \Cvxcol) \coth(\ell_0/2)\bigr).
  \end{align*}
  In particular, we can put
  \begin{equation}
    \label{eq:def.delta}
    \delta_\eps:= \eps^{1/2}
    \max
    \Bigl\{%
      \Bigl(\Cvx + \frac{m\eps}{\lambdaEd}\Bigr) \CGaffney, %
      \bigl(\Cisoper + \Cvxcol) \coth(\ell_0/2)\bigr)
    \Bigr\}^{1/2}
  \end{equation}
  which concludes the proof.
\end{proof}

\subsection{Proof of the remaining results}
\label{ssec:rest.proofs}


\begin{proof}[Proof of \Thm{main-b}]
  We define now the identification operators for \Thm{main-b}.  Note
  that for a shrinking family of embedded graph-like spaces the
  parameter $\tau$ becomes $\eps\tau$.  For each $e \in E$, we
  define the coordinate transform
  \begin{align*}
    \map {\Phi_e}{\wtXepsed=\wt I_e \times \Yepsed}{\Xepsed=I_e \times \Yepsed},
    \qquad
    \Phi_e(\wt s,y)=\bigl((1-\eps \tau)^{-1} (\wt s-(\eps \tau/2)\ell_e),y \bigr)
  \end{align*}
  if $\ell_e<\infty$ (see also \Fig{notation}), where
  $\wt I_e:=\interval {(\eps \tau/2)\ell_e}{(1-\eps \tau/2)\ell_e}$.
  If $\ell_e=\infty$, we simply set
  $\Phi_e(\wt s,y)=(\wt s-\eps \tau,y)$ and
  $\wt I_e=\interval[open right]{\eps \tau}\infty$.  We define an
  identification operator $\map \wtIdOp {\HSeps}{\wtHSeps}$ by
  \begin{align*}
    (\wtIdOp\UU)_v:=\UU_v
    \qquadtext{and}
    (\wtIdOp\UU)_e:=\UU_e \circ \Phi_e.
  \end{align*}
  Then from integration by substitution we obtain
  $\wtIdOpAdj \WW=(1-\eps\tau)(\WW \circ \Phi_e^{-1})$, and hence
  $(\wtIdOpAdj\wtIdOp)_e=1-\eps \tau$ resp.\
  $(\wtIdOp\wtIdOpAdj)_e=1-\eps \tau$ if $\ell_e<\infty$ and $0$ if
  $\ell_e=\infty$, and finally
  \begin{align*}
    \norm{\id \HSeps-\wtIdOpAdj\wtIdOp}=\eps \tau
    \quadtext{and}
    \norm{\id \wtHSeps-\wtIdOp\wtIdOpAdj}=\eps \tau
  \end{align*}
  if there is an edge $e \in E$ with $\ell_e<\infty$.  For the
  resolvent difference we argue similarly as in \Lem{res.diff}: For
  $\KK \in \HSeps$ and $\LL \in \wtHSeps$, we set
  $\UU=\Reseps^\pm \KK$ and $\WW=\wtReseps^\mp \LL$.  Then we have
  \begin{align*}
    \iprod[\wtHSeps]{(\wtIdOp \Reseps^\pm-\wtReseps^\pm \wtIdOp)\KK}{\LL}
    &=\iprod[\wtHSeps]{\wtIdOp \UU}{\xDirac \WW}
      -\iprod[\wtHSeps]{\wtIdOp \mDirac \UU}{\WW}\\
    &=\bigl(\iprod[\wthseps]{\wtidop \uu}{\xGradAdj \Ww}
      - \iprod[\wtHseps]{\wtIdop \mGrad \uu}{\Ww}\bigr)
      + \bigl(\iprod[\xHs]{\wtIdop \Uu}{\xGrad \ww}
      - \iprod[\wthseps]{\wtidop \mGradAdj \Uu}{\ww} \bigr).
  \end{align*}
  For the first difference, we have
  \begin{align*}
    \iprod[\wthseps]{\wtidop \uu}{\xGradAdj \Ww}
    - \iprod[\wtHseps]{\wtIdop \mGrad \uu}{\Ww}
    &=-\sum_{e \in E}
      \int_{\bd \Xepsed} \uu \; \conj
      \Ww \cdot \normvec{\bd \Xepsed} \dd \bd \Xepsed\\
    &=\sum_{v \in V}
      \int_{\bd \Xepsvx} \uu \; \conj
      \Ww \cdot \normvec{\bd \Xepsed} \dd \bd \Xepsvx\\
    &=\sum_{v \in V} \int_{\Xepsvx} \Div (\uu \conj \Ww) \dd \Xepsed\\
    &=\sum_{v \in V} \int_{\Xepsvx}
      \bigl(\uu \; \conj{\GradAdj \Ww} - \iprod[g] {\Grad \uu} \Ww \bigr)\dd \Xepsed
  \end{align*}
  using integration by parts for the first and the divergence theorem
  for the second last equality.  For the second difference, we have similarly
  \begin{align*}
    \iprod[\wtHseps]{\wtIdop \Uu}{\xGrad \ww}
    - \iprod[\wthseps]{\wtidop \mGradAdj \Uu}{\ww}
    &=\sum_{e \in E}
      \int_{\bd \Xepsed} \conj \ww \; \conj
      \Uu \cdot \normvec{\bd \Xepsed} \dd \bd \Xepsed\\
    &=-\sum_{v \in V}
      \int_{\bd \Xepsvx} \conj \ww \; \Uu \cdot \normvec{\bd \Xepsvx} \dd \bd \Xepsed\\
    &=-\sum_{v \in V} \int_{\Xepsvx} \Div (\conj \ww \Uu) \dd \Xepsed\\
    &=-\sum_{v \in V} \int_{\Xepsvx}
      \bigl(\GradAdj \Uu \; \conj w - \iprod[g] \Uu {\Grad \ww}\bigr)\dd \Xepsed.
  \end{align*}
  Now we estimate
  \begin{align*}
    \bigabssqr{\iprod[\xHS]{(\wtIdOp \Reseps^\pm-\wtReseps^\pm \wtIdOp)\KK}{\LL}}
    &\le \sum_{v \in V}\normsqr[\Lsqr \Xepsvx]{\uu_v}
      \normsqr[\Lsqr \wtXeps] {\wtGradAdj \Ww}
      + \normsqr[\Lsqr {T^*\Xeps}] {\Grad \Uu}
      \sum_{v \in V}\normsqr[\Lsqr {T^*\Xepsvx}]{\Ww_v}\\
    & \qquad + \normsqr[\Lsqr \Xeps] {\GradAdj \Uu}
      \sum_{v \in V}\normsqr[\Lsqr \Xepsvx]{\ww_v}
      + \sum_{v \in V}\normsqr[\Lsqr {T^*\Xepsvx}]{\Uu_v}
      \normsqr[\Lsqr \wtXeps] {\wtGrad \ww}\\
    & \le\Cvx\eps
      \bigl(
      \normsqr[\Sob \Xeps] \uu
      \normsqr[\Lsqr \wtXeps] {\wtGradAdj \Ww}
      +  \normsqr[\Lsqr \Xeps] {\GradAdj \Uu}
      \normsqr[\Sob \Xeps] \ww \bigr)\\
    & \qquad + \Cvx\CGaffney\eps
      \bigl(
      \normsqr[\Lsqr {T^*\Xeps}] {\Grad \Uu}
      \normsqr[\wtHsdomeps] \Ww
      + \normsqr[\Hsdomeps] \Uu
      \normsqr[\Lsqr \wtXeps] {\wtGrad \ww}
      \bigr)\\
    & \le \Cvx (1+\CGaffney) \eps \normsqr[\HSeps] \KK \normsqr[\wtHSeps]\LL
  \end{align*}
  using \Lem{vx.est}, \Cor{kato.ineq} and \Thm{gaffney.scaled}.  In
  particular, we have
  \begin{equation*}
    \normsqr{\wtIdOp \Reseps^\pm-\wtReseps^\pm \wtIdOp}
    \le \Cvx (1+\CGaffney) \eps.
  \end{equation*}
  In particular, we can put here
  \begin{equation}
    \label{eq:def.delta'}
    \delta_\eps':= \eps^{1/2}
    \max
    \Bigl\{%
    \tau, %
    \Cvx (1+\CGaffney)
    \Bigr\}^{1/2}
  \end{equation}
  and the theorem is proven.
\end{proof}

\begin{proof}[Proof of \Cor{main-c}]
  The result follows from \Thms{main-a}{main-b}, and the transitivity
  of quasi-unitary equivalence (see
  e.g.~\cite[Lem.~2.11~(c)]{post-simmer:20}).
\end{proof}

\begin{proof}[Proof of \Cor{main-d}]
  The result follows from \Thm{main-a} resp.\ \Cor{main-c} and
  \Prp{conv.lapl}.
\end{proof}


\appendix
\section{Quasi-unitary equivalence and generalised norm resolvent
  convergence}
\label{app:que}
\setboolean{EpsNotation}{false}

Let us briefly review the concept of quasi-unitary equivalence which
can be used to define a \emph{distance} between two bounded operators
acting in different Hilbert spaces (see~\cite{post-zimmer:pre24} for
more details on this viewpoint).  This distance then gives rise to a
natural convergence.  As the operators under consideration here are
unbounded, we use their resolvents.

Let $\mDirac$ resp.\ $\xDirac$ denote two self-adjoint operators
acting in Hilbert spaces $\mHS$ and $\xHS$.
\begin{definition}[quasi-unitary equivalence]
  \label{def:que}
  Let $\delta>0$.  We say that $\A$ and $\xA$ are
  $\delta$-quasi-unitary equivalent if there is a bounded operator
  $\map J \mHS \xHS$ with $\norm [\Lin{\mHS,\xHS}] J \le 1$ such that
  \begin{subequations}
  \begin{align}
    \label{eq:que.1}
    \norm[\Lin \mHS]{(\id \mHS - \IdOpAdj \IdOp) \mRes^\pm}
    &\le \delta,
    &\norm[\Lin \xHS]{(\id \xHS - \IdOp\IdOpAdj) \xRes^\pm}
    &\le \delta,\\
    \label{eq:que.2}
    \norm[\Lin {\mHS,\xHS}]{\IdOp \mRes^\pm - \xRes^\pm \IdOp},
    &\le \delta
  \end{align}
  where
  \begin{align}
    \label{eq:que.3}
    \mRes^\pm
    &:=(\mDirac \mp \im)^{-1},
    &\xRes^\pm
    &:=(\xDirac \mp \im)^{-1}
  \end{align}
  are the resolvents of $\A$ resp.\ $\xA$.
\end{subequations}
\end{definition}
Here and in the sequel, one has to choose either the upper or the
lower sign in one formula.

We can turn the above concept into a convergence as follows: Let
$\mHS=\HSzero$ and $\xHS=\HSeps$ and $\mDirac=\Diraczero$ and
$\xDirac=\Diraceps$.
\begin{definition}
  \label{def:gnrs}
  We say that $\Diraceps$ converges to $\Diraczero$ in
  \emph{generalised norm resolvent sense} if $\Diraceps$ and
  $\Diraczero$ are $\delta_\eps$-quasi-unitarily equivalent with
  $\delta_\eps \to 0$ as $\eps \to 0$.  We refer to $\delta_\eps$ as
  \emph{convergence rate}.
\end{definition}
This convergence is equivalent with a similar one defined by
Weidmann~\cite[Sec.~9.3]{weidmann:00}, which he also called
``generalised norm resolvent convergence'', cf.~\cite{post-zimmer:22}
for details.

If both spaces $\mHS$ and $\xHS$ split into $0$- and $1$-forms as
in~\eqref{eq:hs.1st} and if $\mDirac$ and $\xDirac$ are of the
form~\eqref{eq:first.order'} then we have the following abstract result:
\begin{proposition}
  \label{prp:conv.lapl}
  Assume that $\mDirac$ and $\xDirac$ are $\delta$-quasi-unitarily
  equivalent such that the identification operator respects the form
  structure, i.e. such that
  \begin{align*}
    \map {\IdOp=\idop \oplus \Idop}{\mHS=\mhs\oplus\mHs}{\xHS=\xhs\oplus\xhs},
    \quad
    \FF=(\ff,\Ff) \mapsto (\idop \ff,\Idop \Ff),
  \end{align*}
  then the corresponding Laplacians $\mLAPL=\mlapl \oplus \mLapl$ and
  $\xLAPL=\xlapl \oplus \xLapl$ are $2\delta$-quasi-unitarily
  equivalent.

\end{proposition}
\begin{proof}
  Note first that
  \begin{equation*}
    \mRes^\pm \mRes^\mp
    =(\mDirac^2+1)^{-1}
    =(\mLAPL+1)^{-1}
  \end{equation*}
  and similarly for $\xLAPL$.  Now, the Laplace resolvent difference
  is
  \begin{align*}
    \norm{\IdOp (\mLAPL+1)^{-1} -(\xLAPL+1)^{-1} \IdOp}
    &=\norm{\IdOp \mRes^\pm \mRes^\mp -\xRes^\pm \xRes^\mp\IdOp}\\
    &\le\norm{(\IdOp \mRes^\pm- \xRes^\pm\IdOp)\mRes^\mp}
      + \norm{\xRes^\pm(\IdOp \mRes^\mp - \xRes^\mp\IdOp)}\\
    &\le\norm{\IdOp \mRes^\pm- \xRes^\pm\IdOp}
      + \norm{\IdOp \mRes^\mp - \xRes^\mp\IdOp}
      \le 2\delta
  \end{align*}
  and, as $\IdOp$ and $\mLAPL$ resp.\ $\xLAPL$ preserve the splitting
  into $0$- and $1$-forms, we also have
  \begin{multline*}
    \norm
    {\IdOp (\mLAPL+1)^{-1} -(\xLAPL+1)^{-1} \IdOp}\\
    =\max \bigl\{
    \norm
    {\idop (\mlapl+1)^{-1} -(\xlapl+1)^{-1} \idop},
      \norm
      {\Idop (\mLapl+1)^{-1} -(\xLapl+1)^{-1} \Idop}
    \bigr\},
  \end{multline*}
  i.e., also $\mlapl$ and $\xlapl$ resp.\ $\mLapl$ and $\xLapl$ are
  $2\delta$-quasi-unitarily equivalent.
\end{proof}
One can also show that other functions of the operator such as the
heat operator or spectral projections sandwiched suitably with $\IdOp$
and $\IdOpAdj$ fulfil similar operator norm estimates as above; for
details we refer to~\cite[Ch.~4]{post:12}
or~\cite[Sec.~2.3]{post-simmer:20}, where weaker versions of spectral
estimates are shown; we conclude the appendix with a brief recall of
the current state of the art in this area.


If the identification operator $\IdOp$ is a (partial) isometry, then
the definition of quasi-unitary equivalence given here is equivalent
with
\begin{align}
  \label{eq:que.1'}
  \tag{\ref{eq:que.1}'}
  \norm[\Lin \mHS]{\mRes^\mp(\id \mHS - \IdOpAdj \IdOp) \mRes^\pm}^{1/2}
  &\le \delta,
  &\norm[\Lin \xHS]{\xRes^\mp(\id \xHS - \IdOp \IdOpAdj) \xRes^\pm}^{1/2}
  &\le \delta
\end{align}
used in~\cite{post-zimmer:pre24}.  A proof relies on the fact that $J$
is a partial isometry ($JJ^*J=J$) if and only if the (squared)
so-called \emph{defect operator} $\id \HS -\IdOpAdj \IdOp$ is an orthogonal
projection (cf.~\cite[Thm.~3.9]{post-zimmer:22}, the explanation for
the need for the new version is given
in~\cite[Rem.~5.2]{post-zimmer:pre24}.

In~\cite{post-zimmer:pre24} we defined different distances between
$\mDirac$ and $\xDirac$ and showed, among other results, that they are
all equivalent and imply in particular that
\begin{equation}
  \label{eq:spec.dist}
  \dHaus\bigl(\spec{(\mDirac \mp \im)^{-1}},\spec{(\xDirac \mp \im)^{-1}}\bigr)
  \le \sqrt 3 \delta
\end{equation}
(cf.~\cite[Thm.~A, Cor.~D]{post-zimmer:pre24}), where
$\dHaus(\Sigma,\wt \Sigma)$ is the usual Hausdorff distance.  There is
a finer version respecting multiplicities of discrete eigenvalues
giving the same upper estimate $\sqrt 3 \delta$; we refer again
to~\cite{post-zimmer:pre24} and references therein for further
details.

%
%
\providecommand{\bysame}{\leavevmode\hbox to3em{\hrulefill}\thinspace}
\providecommand{\MR}{\relax\ifhmode\unskip\space\fi MR }
\providecommand{\MRhref}[2]{%
  \href{http://www.ams.org/mathscinet-getitem?mr=#1}{#2}
}
\providecommand{\href}[2]{#2}


\end{document}